\documentclass[fleqn,usenatbib]{mnras}
\usepackage{newtxtext,newtxmath}
\usepackage[T1]{fontenc}
\usepackage{graphicx}	% Including figure files
\usepackage{booktabs,xltabular}

\usepackage{multirow}
\usepackage{csvsimple}
\usepackage{supertabular}

\newcommand{\Eiso}{$E_{\gamma, \rm iso}$}

\DeclareUnicodeCharacter{2212}{-}

\title[GRB 201015A and soft GRBs]{GRB 201015A and the nature of low-luminosity soft gamma-ray bursts}
\author[M. Patel et al.]{M. Patel,$^{1}$\thanks{email:mp664@leicester.ac.uk}
B. P. Gompertz,$^{2}$
P. T. O'Brien,$^{1}$
G. P. Lamb,$^{1,3}$
R. L. C. Starling,$^{1}$
P. A Evans,$^{1}$
\newauthor 
L. Amati,$^{4}$
A. J. Levan,$^{5}$
M. Nicholl,$^{2,6}$
K. Ackley,$^{7}$
M. J. Dyer,$^{8}$
J. Lyman,$^{7}$
\newauthor 
K. Ulaczyk,$^{7}$
D. Steeghs,$^{7}$
D. K. Galloway,$^{9,10,11}$
V. S. Dhillon,$^{8,17}$
G. Ramsay,$^{12}$
\newauthor 
K. Noysena,$^{13}$
R. Kotak,$^{14}$
R. P. Breton,$^{15}$
L. K. Nuttall,$^{16}$
E. Pall\'e,$^{17}$
D. Pollacco$^{7}$
\\
$^{1}$
School of Physics and Astronomy, University of Leicester, University Road, Leicester LE1 7RH, UK\\
$^{2}$
School of Physics and Astronomy \& Institute for Gravitational Wave Astronomy, University of Birmingham, Birmingham B15 2TT, UK\\
$^{3}$
Astrophysics Research Institute, Liverpool John Moores University, IC2 Liverpool Science Park, 165 Brownlow Hill, Liverpool L3 5RF\\
$^{4}$
INAF - IASF Bologna, via P. Gobetti 101, Bologna, Italy\\
$^{5}$
Department of Astrophysics/IMAPP, Radboud University, P.O. Box 9010, NL-6500 GL Nijmegen, The Netherlands\\
$^{6}$
Astrophysics Research Centre, School of Mathematics and Physics, Queen’s University Belfast, Belfast BT7 1NN, UK\\
$^{7}$
Department of Physics, Warwick University, Coventry CV4 7AL\\
$^{8}$
Department of Physics and Astronomy, University of Sheffield, Sheffield S3 7RH, UK\\
$^{9}$
School of Physics \& Astronomy Monash University, Wellington Rd, Clayton VIC 3800, Australia\\
$^{10}$
OzGRav-Monash, School of Physics and Astronomy, Monash University, Victoria 3800, Australia\\
$^{11}$
Institute for Globally Distributed Open Research and Education (IGDORE)\\
$^{12}$
Armagh Observatory \& Planetarium, College Hill, Armagh BT61 9DB\\
$^{13}$
National Astronomical Research Institute of Thailand\\
$^{14}$
Department of Physics and Astronomy, University of Turku, 20500 Turku, Finland\\
$^{15}$
Department of Physics and Astronomy, University of Manchester, Oxford Rd, Manchester M13 9PL\\
$^{16}$
School of Mathematics \& Physics, University of Portsmouth, University House, Winston Churchill Ave, Portsmouth PO1 2UP\\
$^{17}$
Instituto de Astrofisica de Canarias, E-38205 La Laguna, Tenerife, Spain}

\date{Accepted XXX. Received YYY; in original form ZZZ}

\pubyear{2022}

\begin{document}
\label{firstpage}
\pagerange{\pageref{firstpage}--\pageref{lastpage}}
\maketitle

\begin{abstract}
%\textbf{
GRB 201015A is a peculiarly low luminosity, spectrally soft gamma-ray burst (GRB), with $T_{\rm 90} = 9.8 \pm 3.5$ s (time interval of detection of 90\% of photons from the GRB), and an associated supernova (likely to be type Ic or Ic-BL). GRB 201015A has an isotropic energy \Eiso{}$ = 1.75 ^{+0.60} _{-0.53} \times 10^{50}$ erg, and photon index $\Gamma = 3.00 ^{+0.50} _{-0.42}$ (15--150 keV). It follows the Amati relation, a correlation between \Eiso{} and spectral peak energy $E_{\rm p}$ followed by long GRBs. It appears exceptionally soft based on $\Gamma$, the hardness ratio of HR = $0.47 \pm 0.24$, and low-$E_{\rm p}$, so we have compared it to other GRBs sharing these properties. These events can be explained by shock breakout, poorly collimated jets, and off-axis viewing. Follow-up observations of the afterglow taken in the X-ray, optical, and radio, reveal a surprisingly late flattening in the X-ray from $t = (2.61 \pm 1.27)\times 10^4$~s to $t = 1.67 ^{+1.14} _{-0.65} \times 10^6$~s. We fit the data to closure relations describing the synchrotron emission, finding the electron spectral index to be $p = 2.42 ^{+0.44} _{-0.30}$, and evidence of late-time energy injection with coefficient $q = 0.24 ^{+0.24} _{-0.18}$. The jet half opening angle lower limit ($\theta_{j} \ge 16^{\circ}$) is inferred from the non-detection of a jet break. The launch of SVOM and Einstein Probe in 2023, should enable detection of more low luminosity events like this, providing a fuller picture of the variety of GRBs.
\end{abstract}


\section{Introduction}
Several thousand gamma-ray bursts (GRBs) have been detected since their first identification more than 50 years ago \citet{Klebesadel73}. These bursts of gamma-rays are detected by satellites such as the \emph{Neil Gehrels Swift Observatory} \citep{Gehrels04} or \emph{Fermi} \citep{Meegan09}. For the majority of these events, information is only available in the $\gamma-$ray bands, typically in the range from tens of keV to a few MeV, providing both temporal and spectral information \citep{Piran04}. On the basis of these data alone the population is highly varied, but the duration of the bursts is clearly bimodal, splitting the long and short GRBs at a boundary at $T_{\rm 90} \sim 2$~s \citep{Kouveliotou93}, where $T_{\rm 90}$ is the duration in which 90\% of the photons from the GRB are detected \citep{Koshut95}. The short GRB population typically has harder spectra, meaning there is a larger proportion of higher energy photons to lower-energy photons, compared to long GRBs \citep{Zhang16}. 

% new para here where you start talking about MW information
Intensive multi-wavelength observations of GRBs over the past 30 years have
revealed that these two populations arise from distinct progenitors \citep[e.g.][]{Chevalier99, Levan16}.  
The long GRBs (LGRBs), lasting typically from a few seconds up to several minutes or longer, arise from star forming galaxies \citep{Fruchter06}, and photometric and spectroscopic
monitoring has revealed that they are created in the core-collapse of massive, rapidly spinning, likely low metallicity stars creating a supernova type I b/c counterpart to the GRB \citep{Hjorth03,Stanek03,Levan18}. This event is also referred to as a collapsar. The first evidence of this origin of long GRBs came from the direct association of GRB980425 with SN1998bw, a peculiar type Ib/c SN \citep{Galama99}.

In contrast, the short GRBs (SGRBs), lasting from few hundreds of ms up to 2 s \citep{Kouveliotou93}, are found in galaxies of all ages, sometimes at large distances from their hosts \citep{Fong22}. The identification of possible kilonovae \citep{Tanvir13, Troja17b, Lamb19b}, powered by radioactive decays
of r-process elements, and ultimately the near-simultaneous detection of a short GRB with the gravitational wave detected merger GW170817 secured their
origin in the mergers of compact objects \citep{Abbott17a}. Although, short GRBs are almost always viewed on-axis, GRB 170817A was viewed at $\leq$ 36 degrees from the jet axis \citep{Abbott17a}, and GRB 150101B, was significantly under-luminous and later shown to be viewed off-axis by 13 degrees \citep{Troja18}.

For both the collapsar or compact object merger scenarios, the central engine (an accreting black hole or neutron star) launches relativistic jets of material \citep[e.g.][]{Kluzniak98, Lei13, Lu15}. Gamma-rays are produced via either self-interactions of this material \citep{Sari99, Meszaros02} or the dissipation of magnetic fields \citep{Beniamini17}. As the jets plough into the circumstellar environment they are decelerated and shock fronts interact with the surrounding material, forming a broadband synchrotron `afterglow' \citep{Gao13b, Iyyani16}.

There are suggestions of additional categorisations of GRBs, for example
a separate ``intermediate" duration population lasting 2 -- 5 s \citep[e.g.][]{Mukherjee98, Tunnicliffe12}, low-luminosity GRBs with luminosity $L < 10^{49}$ erg s$^{-1}$ \citep{Liang07, Virgili08}, or events which are ``ultra-long" \citep{Levan14}. However, the reality of these populations, and if they represent distinct physical 
processes remains unclear. We recognise that there is a lot of variation within the GRB population. They range in duration from milliseconds to hours, have spectral peaks ranging from the keV to MeV range, and isotropic energies ranging from $10^{46}$ -- $10^{54}$ erg \citep{Virgili08, Levan16}. Some have very smooth light curves, others demonstrate pronounced variability \citep{Zhang16}. It is therefore quite plausible that additional mechanisms are present within the observed populations \citep{Nousek06}. 

Indeed, it is striking that there are a small number of apparently long GRBs which do not exhibit supernova signatures, which are suggested
to arise from mergers such as GRB211211A, which has an observed kilonova counterpart \citep{Rastinejad22}. These SGRBs are detected as long GRBs based on their $T_{\rm 90}$ in the $\gamma$-ray band, but have multiple spikes with extended emission (EE-SGRBs) \citep[e.g.][]{Gehrels06,Zhang20, Gompertz22}. There are also discoveries of short-duration GRBs with a core-collapse supernova association \citep[e.g.][]{Amati21, Ahumada21, Zhang21}.
A re-analysis of the prompt GRB emission properties using a machine learning based approach is largely successful at splitting GRBs into the two merger and collapsar groups, however, some GRBs are incorrectly classified, and some fail to be robustly classified \citep{Jespersen20}.
%\rlcs{add a ref or two for each 'grouping' suggested, eg for ultralong can use Levan et al 2014}
%\bg{I added a few citations, including the Levan one, but more are probably needed}

To understand the nature of GRBs therefore continues to require further observations, in particular of bursts which appear to defy ready classification within a single scheme, for example long-GRBs without associated supernovae or in older galaxies; GRBs which appear under or over-luminous, or bursts which touch on the extremes of duration, spectrum, variability or other key indicators. 

Here we consider the case of GRB 201015A, a GRB close to the long-short GRB divide, exhibiting a short peak and extended emission morphology in the prompt emission light curve \citep{GCN28658}. Strikingly, it was also extremely spectrally soft, much softer than the majority of short, or even long-GRBs \citep{Lien16}. The detection of a supernova confirms this GRB to have a collapsar progenitor \citep{GCN29306}. 

This GRB has been of interest due to the possible very high energy (VHE) detection at TeV energies \citep{GCN28659}.
The MAGIC collaboration reported a 3.5 sigma detection of the GRB 201015A in TeV energies beginning 33s after the trigger \citep{Suda21}. This makes GRB 201015A the fifth burst to be detected in very high energy (VHE), giving possible further evidence of Synchrotron self-Compton (SSC) emission \citep{Nava18}. This is the process of Synchrotron photons - produced from electrons accelerated in a magnetic field - scattering off electrons making them more energetic. The other GRBs detected in VHE by the MAGIC and H.E.S.S. collaborations are GRB 190114C, GRB 180720B, GRB 190829A, and GRB 201216C. This sample already includes a variety of both high and low prompt energy GRBs with GRB 190114C having \Eiso{}$= 3 \times 10^{53}$ erg and GRB 190829A having \Eiso{}$ = 2 \times 10^{50}$ erg \citep{Berti22}. GRB 201015A adds another low luminosity GRB to the sample.

The afterglow observations in radio by the VLBI telescope as well as afterglow measurements in other wavebands have been studied. \citet{Giarratana22} compares the GRB 201015A to other VHE bursts, and uses afterglow models to suggest this is an on-axis GRB expanding into a homogeneous ISM-like medium. In this paper, we look into the initial energetics of the burst, comparing the isotropic energy, \Eiso{}, and spectral peak energy $E_{\rm p}$, to the Amati relation for long GRBs \citep{Amati06}. We identify and compare a sample of similar bursts in terms of the soft prompt spectrum and low-luminosity and low-Epeak. We have also collated multi-wavelength information on the afterglow of GRB 201015A and related this to models to explain an observed break in the X-ray afterglow. 
%, giving evidence of late-time energy injection at $\sim 2 \times 10^{4}$ s. We find that the sample of low-luminosity and low-Epeak bursts discovered so far exhibit a range of conditions causing these observational properties. They tend to have some features in common such as shorter duration than most long bursts and a softer prompt spectrum.
%\rlcs{Don't give away the results just yet! Just say here what you will investigate in the rest of the paper.}
%\bg{seconded}

In Section~\ref{sec:obs}, we present the data we have collected of the prompt and afterglow emission of GRB 201015A, and Section~\ref{section:analysis} goes through the fitting of models to the afterglow and SN, as well as predicting the duration of the X-ray plateau, and minimum jet opening angle. We then compare GRB 201015A to other spectrally soft GRBs in Section~\ref{sec:softpop}, and discuss our findings in Section~\ref{sec:Discuss}, and provide conclusions in Section~\ref{sec:Conclusion}. The optical data is fairly extensive, so it is presented in the Appendices. The findings presented in this paper are quoted with 1-sigma confidence regions. A flat $\Lambda CDM$ cosmology with $\Omega_{m} = 0.3$, $\Omega_{\Lambda} = 0.7$, $H_0 = 65$ km s$^{-1}$ Mpc$^{-1}$ has been assumed for this work. In this paper, we define the spectral index, $\beta$, as $F(\nu)_{\nu} \propto \nu^{-\beta}$ and similarly for the temporal power-law relation $F(\nu)_{t} \propto t^{-\alpha}$. The equation describing the relation between the photon index, $\Gamma$, and spectral index, $\beta$, is: $\Gamma = \beta + 1$.
%\bg{Add a paragraph here outlining the sections and defining any conventions, e.g. cosmology, sign conventions for power laws, quoted confidence intervals.}

\section{Observations}\label{sec:obs}
\subsection{Discovery of GRB 201015A}

Here we present the observations detailed in the public Gamma-ray Burst Coordinates Network (GCN). 
The \emph{Swift} Burst Alert Telescope \citep[BAT;][]{Barthelmy05a} triggered on GRB 201015A at 22:50:13 UT (hereafter $T_0$) \citep{GCN28632}. The refined BAT analysis reveals the position of this source to be RA = 354.342, DEC = 53.393 \citep{GCN28658}. The BAT light curve shows a short initial spike of overlapping pulses that lasts for $\sim 1$s, followed by a tail of extended emission that lasts until $\sim10$s, exhibiting a similar morphology to EE-GRBs \citep{Norris10}. The prompt emission in the 15-350~keV range has a duration of $T_{\rm 90}  = 9.78 \pm 3.47$s \citep{GCN28658}. The time-averaged BAT spectrum (from $T+0.02$ to $T+10.35$ s) is best fit by a power-law model with a photon index of $\Gamma = 3.03 \pm 0.68$ \citep{GCN28658}.
%The fluence in the $15-150$~keV band is $(2.0 \pm 0.6) \times 10^{-7}$~erg~cm$^{-2}$ \citep{GCN28658}. %At the reported redshift of the GRB of $z = 0.426$ \citep{GCN28649, GCN28661}, 

The \emph{Fermi} Gamma-ray Burst Monitor \citep[GBM;][]{Meegan09} detected a weak, sub-threshold event with a duration of $\sim 1$s \citep{GCN28663}. The spectrum is adequately fit by a Band function \citep{Band93} with a peak energy of $E_{\rm p} = 14 \pm 6$~keV, a fixed low-energy index of $\beta_1 = 1$, and a high-energy index of $\beta_2 = 2.40 \pm 0.21$. This model yields a 10-1000~keV fluence of $(2.25 \pm 0.38) \times 10^{-7}$~erg~cm$^{-2}$ \citep{GCN28663}, resulting in \Eiso{} $ = (1.1 \pm 0.2) \times 10^{50}$~erg. %\bg{If the above is all quoted from GCNs then it should go in the introduction}

Due to an observing constraint, \emph{Swift} was unable to slew to GRB 201015A until T$_{0}$+$51.6$ minutes. Once on target, the X-ray Telescope \citep[XRT;][]{Burrows05} detected a fading, uncatalogued X-ray source within the BAT error circle. The combined XRT-UVOT observations reveal a source at RA(J2000) = 23:37:16.43, Dec(J2000) = +53:24:57.5 with an uncertainty of 1.9 arcseconds (radius, 90 per cent containment) \citep{GCN28647,Goad07, Evans09}.
The XRT (0.3 -- 10 keV) afterglow light curve, between $0.03$ and $21$ days after the trigger, decays with a power-law index of $\alpha_x = 1.80^{+0.22}_{-0.20}$, according to the live XRT GRB catalogue\footnote{\url{https://www.swift.ac.uk/xrt_live_cat/01000452/}} \citep{Evans09}.

The MAGIC telescope's 3.5$\sigma$ detection of the source came from almost 4 hours of observation starting at $T_0 + 33s$ finding evidence of $> 140 GeV$ emission from GRB 201015A \citep{Suda21}.

An optical transient coincident with the BAT error circle was first reported by MASTER \citep{Lipunov10} 88s after the trigger \citep{Lipunov20} and prior to the XRT detections. This was confirmed to be the afterglow by the Nordic Optical Telescope \citep[NOT;][]{Malesani20} and the Gravitational-wave Optical Transient Observer \citep[GOTO;][]{Ackley20}. An uncatalogued host galaxy with magnitude $r=22.9 \pm 0.2$ was identified coincident with the GRB position, separated by 2.3'' \citep{GCN28676, GCN29306}. Optical observations by GTC/OSIRIS started at $T_0+5.28$ hours providing a spectrum from 3700 to 7800 \AA{}. A redshift of $z=0.426$ was calculated from [OIII], [OII], and H-$\beta$ emission lines identified above the continuum \citep{GCN28649}.

Radio observations were taken by the VLA at a central frequency of 6 GHz 1.41 days after the trigger showing a flux density of $\sim 0.13$ mJy \citep{GCN28688}. Following this the e-Merlin telescope observed the transient at a central frequency of 1.5 GHz at 19 and 23 days after the trigger. The measured flux densities were $(2.14 \pm 0.25) \times 10^{-4}$ Jy and $(2.56 \pm 0.27) \times 10^{-4}$ Jy respectively \citep{GCN28945}. Both sources were found to be at a consistent position with the optical afterglow.
 
%optical supernova obs and redshift \citep{GCN28649}

\subsection{Gamma-rays}\label{section:gamma}

We downloaded the BAT spectral files from the \emph{Swift}/BAT catalogue \citep{Lien16} to perform an independent analysis using {\sc xspec} v12.11.1. 
%The data were processed with the BAT pipeline {\sc batgrbproduct} using HEAsoft v6.28. \textcolor{red}{??}
%We extracted a spectrum in the interval $-0.5 \leq T_0 \leq 2$s. The spectrum is well fit by a power-law model with a photon index of $\Gamma = 3.64 \pm 0.47$ (90 per cent confidence interval. Reduced $\chi^2 = 0.84$ for 56 degrees of freedom). We note that the spectrum is equally well fit by a blackbody with $kT = 3.96 \pm 0.51$~keV (90 per cent confidence interval. Reduced $\chi_{\nu}^2 = 0.87$ for 56 degrees of freedom).
The time-averaged total BAT spectrum is referred to as the BAT $T_{\rm 100}$ in this paper.
 We fit a singe power-law to the BAT $T_{\rm 100}$ spectrum with duration 10.32 s in the 15-150 keV energy bands using {\sc xspec}, fitting with chi-squared statistics. The result is a photon index of $\Gamma = 3.00^{+0.50}_{-0.42}$ with normalisation $K = 200_{-200}^{+744}$ photons keV$^{-1}$ cm$^{-2}$ s$^{-1}$ at 1 keV. The fit has a reduced chi-squared value of $ \chi_{\nu}^{2} = 1.03 $ with $56$ degrees of freedom.
The measured photon index is unusually soft, particularly in the BAT band. \citet{Lien16} showed that the photon indices measured for single power law fits to BAT data of SGRBs are distributed around $\Gamma \approx 1.5 \pm 0.5$, and for LGRBs around $\Gamma \approx 2.0 \pm 0.5$. In fact, only three other GRBs in the sample of \citet{Lien16} have measured photon indices $\Gamma > 3$ with constrained confidence regions. These GRBs are: 050416A, 080520 and 140622A, and we have analysed them in this paper.
 
In order to compare the energetics of GRB 201015A to other GRBs, we calculate the spectral peak energy, $E_{\rm peak}$, and isotropic equivalent energy, \Eiso{}. The time-averaged BAT spectrum (with exposure time $T_{\rm 100} = 10.32 s$) in the 15-150 keV band was used to fit both a cut-off power-law ({\sc cutoffpl}) and Band function ({\sc grbm}) \citep{Band93} using {\sc xspec} to determine the spectral peak. For these two spectral models, it was difficult to constrain the 1 sigma confidence region on some of the parameters including the spectral break. Using an f-test to determine the best fit we found that the simple power-law described above provided a better fit for this GRB than a model with additional parameters.
%For this GRB the spectral peak energy is close to the low energy threshold of BAT so we fixed the low-energy power-law index, $\alpha$, to -1 based on the canonical values \citep{Lien16}, as done by the Fermi team \citep{GCN28663}. This helped constrain the upper 1 sigma confidence region for the characteristic energy, $E_0$, but the lower error could not be constrained. 

%We find the best fit to be the Band function with the following parameters: $\alpha = -0.3$, $\beta = -3.0_{-0.93}^{+0.49}$, $E_0 = 10.4_{-10.4}^{+2.7}$ keV, and normalisation $K = 0.104_{-0.014}^{+0.027}$.
A summary of the fitting can be found in the Table~\ref{tab:BAT_fit}.
After finding a suitable fit, the normalisation parameter was frozen, and the {\sc cflux} component was added to the model to determine the flux. 

\begin{table}
    \centering
    \begin{tabular}{|c|c|c|c|}
    \hline \hline
         Spectrum & Model & Parameters & $\chi_{\nu}^{2}$ \\
         \hline
         %BAT &power-law& $\Gamma = 2.82^{+0.42}_{-0.36}$ & 0.95\\
         %$T_{\rm 90}$& & $K = 108\pm 137$ cts keV$^{-1}$ cm$^{-2}$ s$^{-1}$ &  \\
         %\hline
         BAT  &power-law& $\Gamma = 3.00_{-0.42} ^{+0.50}$ &  1.03 \\
         $T_{\rm 100}$& & $K = 200_{-200}^{+744}$ cts keV$^{-1}$ cm$^{-2}$ s$^{-1}$ & \\
         \hline \hline
    \end{tabular}
    \caption{Table of spectral model parameters resulting from {\sc xspec} fitting of BAT spectra of GRB 201015A.}
    \label{tab:BAT_fit}
\end{table}

The intrinsic spectral peak energy is calculated using the characteristic energy $E_0$, power-law index $\beta$, and redshift $z$, with the following formula.
\begin{equation}
\label{eq:Ep}
    E_{\rm peak,i} = E_0 (2-\beta) (1+z)
\end{equation}
For this GRB, however, we have not observed the spectral peak energy, so it is assumed to be below the lower-limit of the detector bandpass (15 keV), which is consistent with the result of $14 \pm 6$ keV from the Band function fit to the Fermi spectrum \citep{GCN28663}. Correcting for redshift results in $E_{\rm peak,i} < 21.39$ keV.

The isotropic equivalent energy of the burst was calculated using:
\begin{equation}
    E_{\gamma,\rm iso} = F_{\rm d} T_{\rm 100} \frac{F_{\rm iso}}{F_{\rm d}} \frac{4 \pi D_{\rm L}}{1+z}
\end{equation}
where $F_d$ is the flux in the detector energy band, $F_{iso}$ is the flux in 10 -- 10,000 keV in the rest frame of the GRB, $T_{\rm 100}$ is the duration of the burst, $D_L$ is the luminosity distance, $z$ is the redshift. We find that \Eiso{}$ = 1.75 ^{+0.60}_{-0.53} \times 10^{50}$ erg for this burst. This is a low \Eiso{} value for a GRB considering most GRBs fall in the range of $10^{52-54}$ ergs \citep{Levan16}. The $E_{\rm p}$ value is also lower than most GRBs which have a typical $E_{\rm p} \sim 250$ keV \citep{Soderberg04}. Previously, events with $E_{\rm p} < 30$ keV were classified as X-ray flares (XRF) \citep{Zhang20b}.

\subsection{X-rays}\label{Xrays}

 %However, we find that the data point centred around $1.7 \times 10^4$\,s after trigger can be rebinned into two individual snapshots with all of the counts in the first bin. Modifying the light curve in this way, we find $\alpha_x = 1.51 \pm 0.14$ for the early ($t < 0.8$ days) X-ray temporal evolution (Reduced $\chi^2 = 1.09$ for 8 degrees of freedom).
The early time-averaged XRT spectrum ($T_0 + 3217$ to $22019$ s) produced by the \emph{Swift} Burst Analyser\footnote{\url{https://www.swift.ac.uk/burst_analyser/}} \citep{Evans10} using HEASoft v6.29 is well fitted within {\sc xspec} \citep{Arnaud96} . We used a composite model comprised of a power-law $\times$ {\sc tbabs} \citep{Wilms00} $\times$ {\sc ztbabs}. This yields a photon index of $\Gamma = 2.16 _{-0.23}^{+0.24}$ and an intrinsic absorption column of $n_{H,z} = (2.1 _{-1.9}^{+2.4}) \times 10^{21}$~cm$^{-2}$ at the redshift of the GRB ($z=0.426$), with normalisation $(5.7_{-1.2}^{+1.6}) \times 10^{-4}$ photons keV$^{-1}$ cm$^{-2}$ s$^{-1}$ at 1 keV (C-stat = 132.85 for 132 degrees of freedom). Fitting was performed using Cash statistics \citep{Cash79}, and the Galactic absorption column was fixed to $n_H = 3.60 \times 10^{21}$~cm$^{-2}$ \citep{willingale13}. We added a blackbody component to the model to check for thermal emission, but this model did not provide a fit as good as the simple power-law with absorption. The results of the best spectral fit are provided in Table~\ref{tab:XRT_fit}.

\begin{table*}
    \centering
    \begin{tabular}{|c|c|c|c|c|}
    \hline \hline
         spectrum & model & parameters & C-stat&degrees of freedom \\
         \hline
         XRT  & power-law $\times$ {\sc tbabs} $\times$ {\sc ztbabs} & $\Gamma = 2.16^{+0.24}_{-0.23}$ & 132.85 & 132\\
         $T_0 + (3217 - 22019) s$& & $n_{H} = 3.6 \times 10^{21}$ & & \\
         & & $z = 0.426$ &  &\\
         & & $n_{H,z} = (2.1^{+2.4}_{-1.9} )\times 10^{21}$ & & \\
         & & $K = (5.7^{+1.6}_{-1.2} )\times 10^{-4}$ photons keV$^{-1}$ cm$^{-2}$ s$^{-1}$&  &\\

         \hline \hline
    \end{tabular}
    \caption{Table of spectral model parameters resulting from {\sc xspec} fitting of the XRT time-averaged spectrum of GRB 201015A.}
    \label{tab:XRT_fit}
\end{table*}

We triggered target of opportunity (ToO) observations with the \emph{Chandra} X-ray Observatory under proposal ID 22400511 (PI: Gompertz). We obtained two epochs of observations with the ACIS-S instrument in Very Faint (VF) mode on the 24$^{\rm th}$ and 29$^{\rm th}$ of October. The exposure times were 35 ks and 45 ks, respectively. The afterglow is clearly detected in both epochs, with $0.5$ -- $7$\,keV count rates of $(4.07 \pm 0.38)\times 10^{-3}$\,cts/s $8.4$ days after trigger and $(3.11 \pm 0.29)\times 10^{-3}$\,cts/s $13.6$ days after trigger.

Data were analysed using {\sc xspec v12.11.1}. In order to account for the possibility of a spectral change during the large gap in coverage between the early XRT data and our first \emph{Chandra} observation, we performed a simultaneous spectral fit of the two \emph{Chandra} epochs and the late \emph{Swift} epoch. This is the same data processing method as used in \citet{Giarratana22}. The simultaneous spectral model is the same as for the early XRT data (i.e. power-law $\times$ {\sc tbabs} $\times$ {\sc ztbabs}), with absorption fixed to the previous values. Our best fit is $\Gamma = 2.10 \pm 0.13$, indicating that the spectrum has not changed significantly.

The full X-ray light curve is shown in Table~\ref{tab:xrays}. XRT data have been absorption corrected using the ratio of counts-to-flux unabsorbed over counts-to-flux observed reported in the time-averaged spectral fit on the UKSSDC. For the \emph{Chandra} data, we extract the unabsorbed $0.3$ -- $10$\,keV fluxes using the {\sc xspec} routine {\sc cflux} and the model fit described previously. Fluxes are then converted to flux densities using \citep[cf.][]{Gehrels08}

\begin{equation}
    F_{\nu, x} = 4.13\times10^{11} \frac{F_x (2-\Gamma)E_0^{1-\Gamma}}{E_2^{2-\Gamma}-E_1^{2-\Gamma}} \mbox{ $\mu$Jy.}
\end{equation}

$E_0$ is the flux density energy (we set 1~keV). $E_1$ and $E_2$ are the lower and upper bounds of the flux bandpass, respectively. 

The resulting light curve is well fit ($\chi_{\nu}^2 = 0.78$ with 10 degrees of freedom) with a broken power-law model with indices $\alpha_1 = 1.53 \pm 0.14$ and $\alpha_2 = 0.48 \pm 0.12$ either side of a break at $t_b = (2.61 \pm 1.27) \times 10^4$s (Figure~\ref{fig:Afterglow_lc}). This differs from the fitting of the X-ray data in \citet{Giarratana22} where this data was approximated to a simple power-law. The broken-power law model for the X-ray afterglow implying a plateau phase affects the closure relation model which describes the synchrotron emission being observed. This long-lasting plateau at late-time is unexpected since it is ongoing until at least $\sim 10^6$~s, as \citet{Tang19} found the plateau end time falls between $(0.9-10)\times10^3$s for 50\% of GRBs with a plateau in the X-ray. We consider the causes for this plateau in Section \ref{section:CR} by analysing the closure relations that best fit this afterglow.

%Jet opening angle constraints, added by Rhaana in case of use\\
%The X-ray light curve does not show any indication of a jet-like break, which is quite common \citep{Curran08}. However, since we have a late epoch measurement at 1.76 $\times$ 10$^{6}$\,s (Table \ref{tab:xrays}) we can provide some reasonable constraints on the jet opening angle \citep{Frail01,Starling08}. 
%Taking the isotropic energy to be  $E_{\gamma,iso} = 1.63^{+0.66}_{-0.52} 10^{50}$ erg, typical values for efficiency $\eta_{\gamma} = 0.2$ and density $n = 0.1$ cm$^{-3}$, and assuming a uniform jet within the standard fireball model, we find that $\theta_j \ge 30^{\circ}$. In turn, this would lead to a collimation-corrected energy of $E_{\gamma} \sim \Eiso{}(1 - \cos \theta_j) = 2.18^{+0.88}_{-0.70} \times 10^{49}$ erg. With no evidence of a jet break or achromatic break, there is little support for this being an off-axis view, making the off-axis scenario an unlikely explanation for the steep-to-shallow transition \citep{Ryan15, Lamb21}.\\

%\rlcs{'which is quite common' is unclear - break is common or no break is common? I know you mean the latter, but think you could rephrase to 'in line with the majority of GRBs' or 'and we note that jet breaks can be very difficult to observe'. The final sentence has a couple of repeat words so could also be rephrased. I think we want to take care we don't oversimplify, and acknowledge that the jet structure can be complex in addition to off-axis and additional jet/cocoon components which we do not treat here.}

\renewcommand{\arraystretch}{1.1}
\begin{table}
\begin{tabular}{|c|c|c|c|}
\hline \hline
Mean Time & Exposure (s) & Flux Density & Telescope \\
since trigger (s) &  & at 1 keV ($\mu$Jy) & \\
\hline
3315.92 & 167.99 & $1.42 \pm 0.31$ & XRT \\
3478.46 & 193.06 & $1.18 \pm 0.27$ & XRT \\
3695.51 & 250.73 & $0.90 \pm 0.20$ & XRT \\
3995.15 & 335.97 & $1.01 \pm 0.19$ & XRT \\
4529.49 & 271.96 & $0.78 \pm 0.21$ & XRT \\
4758.91 & 300.88 & $0.78 \pm 0.17$ & XRT \\
9997.85 & 260.76 & $0.66 \pm 0.17$ & XRT \\
10627.34 & 882.58 & $0.19 \pm 0.05$ & XRT \\
13568.17 & 11036.42 & $0.14 \pm 0.03$ & XRT \\
56362.31 & 1795.23 & $(3.45_{-1.43}^{+1.94})\times 10^{-2}$ & XRT \\
66939.88 & 1384.03 & $< 0.12$ & XRT \\
725760.00 & 28800.00 & $(1.56 \pm 0.07)\times 10^{-2}$ & ACIS \\
1179360.00 & 43200.00 & $(1.36 \pm 0.05)\times 10^{-2}$ & ACIS \\
1760376.71 & 322497.18 & $(8.48_{-2.72}^{+3.27})\times 10^{-3}$ & XRT \\
\hline \hline
\end{tabular}
\caption{X-ray observations of GRB 201015A with \emph{Swift}-XRT and \emph{Chandra}-ACIS.}
\label{tab:xrays}
\end{table}

\subsection{Optical}

The GOTO telescope \cite{Steeghs22} began observations of the target 51 mins after the trigger with 4$\times$90s exposures using the wide L-band filter
(400-700 nm). The source was identified and photometrically calibrated in the following manner. A first pass at source detection is made using {\sc SExtractor} \citep{Bertin96} to identify source positions and preliminary instrumental magnitudes. From the catalog positions, an initial astrometric solution is generated using{\sc astrometry.net} \citep{Lang10}. This solution is further refined if necessary by cross-matching the solved positions against the ATLAS-REFCAT2 \citep{Tonry18} astrometric catalog. Any further refinements to the SIP (Simple Imaging Polynomial) distortion parameters of the
WCS (World Coordinate System) solution is completed using a custom package\footnote{\url{https://github.com/GOTO-OBS/goto-astromtools}}.  Using cross-calibration against the same ATLAS-REFCAT2 catalog and using magnitude zeropoints calibrated against the AAVSO Photometric All-Sky Survey (APASS) survey\footnote{\url{http://www.aavso.org/apass}}, an equivalent APASS  $g'$-band magnitude of the optical afterglow was found to be $g=20.54\pm 0.21$.  %\bg{One of the GOTO team could check this paragraph. We need confidence intervals on the magnitude and I don't remember who did it (wasn't me) or if it's from the GCN perhaps? I don't think the GOTO photometry was done with SEP. That was what I used for the LT photometry. Agree with Rhaana - we don't need to discuss the field here.}
%\rlcs{this is too early to be affected by host light - we should say this, and in the previous paragraph could give the host mag from Rastinejad. Is is corrected for any extinction (Galactic is known so it could be, while host probably not until you do the fitting)? The mention here of the complicated field is premature - we haven't said anything about what that means yet. I would just leave that out - if you want to justify the 3 pixel radius (is that small? we don't know the pixel scale to judge) then say to avoid light from nearby sources.}

We triggered target of opportunity (ToO) observations with the Liverpool Telescope \citep[LT;][]{Steele04} under program PL20B21 (PI: Gompertz). We obtained four epochs of observations with the IO:O instrument using the $r'$ and $i'$ filters \citep{Fukugita96}, beginning on the night of the 16$^{\rm th}$ of October ($\sim$1 day after trigger). The individual images are aligned using {\sc spalipy}\footnote{\url{https://pypi.org/project/spalipy/}} and stacked. Photometry is performed with {\sc sep}\footnote{\url{https://sep.readthedocs.io/en/v1.0.x/index.html}} \citep{Bertin96,Barbary16}. Magnitudes are extracted with a 3 pixel aperture radius, which minimises the confidence interval while avoiding unwanted light from the complicated field. Photometric zero points were computed using nearby field stars in the Pan-STARRS catalog \citep{Chambers16}.

The optical data collected from all the GCNs is shown in the Appendices. These data were then corrected for extinction based on the position of the afterglow and the waveband of the observation using the IRSA Galactic Dust Reddening and Extinction tool\footnote{\url{https://irsa.ipac.caltech.edu/applications/DUST/}}. The magnitudes from the various filters were converted to AB magnitudes and flux density (Jy) using the zero points from \citet{Frei1994}. 

We fit a broken power-law to the afterglow light curve up to $3 \times 10^{5}$ s. We find that the peak is at $t_{d} = 226 \pm 26$ s which is the deceleration time. After the peak, the power-law decays at $\alpha = 0.94 \pm 0.16$. The data after 1 day are fit to the analytic supernova model from \citep{Bazin11} described in Section~\ref{sec:SN}. 

Limited spectral information was available from this data, but we used the simultaneous observations in g', r', and i' bands by the NUTTela-TAO Burst Simultaneous Three-Channel Imager (BSTI) instrument, to infer the spectral index over the time period $\sim 150 - 600$ s. We fit a power-law to the extinction corrected magnitudes against central wavelength of the filters for each epoch with multiple filter observations. We then calculate the average spectral index with weighting of $1-(\sigma/\rm sum(\sigma))$ resulting in average spectral index of $0.75 \pm 0.39$ in the early optical data.
%The observational field is quite complicated; in addition to the underlying host galaxy, there is a galaxy identified in Pan-STARRS \citep{Kaiser02} imaging directly adjacent to the afterglow location. \bg{Should add details of the photometry here (SEP), the extracted magnitudes and perhaps an image like Rhaana suggested.}
%Rhaana: do you think more detail is needed here? eg how close is he nearby galaxy and what mag cf the host? Do we need to show the image to illustrate?

\subsection{Radio}
In Table~\ref{tab:radio}, we have tabulated the radio (1.5 GHz) observations by e-Merlin taken on 3$^{\rm rd}$ and 7$^{\rm th}$ November 2020 presented in the GCN notices \citep{GCN28945}. We fit a simple power-law to the flux densities to determine the decay of the radio light curve: $\alpha \sim -0.94$. Suitable errors could not be provided for this fit due to the lack of observations. 

\begin{table}
    \centering
    \begin{tabular}{|c|c|c|c|}
    \hline \hline
         Time (s) & Flux density (Jy) & Frequency & Telescope\\
         \hline
         $1.642 \times 10^{6}$ & $(2.14 \pm 0.25) \times 10^4$ & $1.5 \times 10^{9}$ & e-Merlin\\
         $1.987 \times 10^{6}$ & $(2.56 \pm 0.27) \times 10^4$ & $1.5 \times 10^{9}$ & e-Merlin\\
         \hline \hline
    \end{tabular}
    \caption{Tabulated values of the radio observations of GRB 201015A reported in \citet{GCN28945}}
    \label{tab:radio}
\end{table}

\section{Analysis}\label{section:analysis}

\subsection{Closure Relations}\label{section:CR}
The relations between the spectral index $\beta$ and the temporal index $\alpha$ based on the synchrotron external shock model, are termed the closure relations \citep{Sari98,Zhang04}.
As the GRB ejecta jet propagates forward, it collides with the circumburst medium creating forward and reverse shock waves, resulting in further shocks. 
This relation arises from the synchrotron radiation emitted by the accelerated electrons in the magnetic field.
The afterglow is described as $F_{\nu} \propto  \nu^{-\beta}t^{-\alpha}$
The electron spectral index $p$ determines the parameters of the closure relation.

\begin{figure*}
\begin{center}
\includegraphics[width=\textwidth]{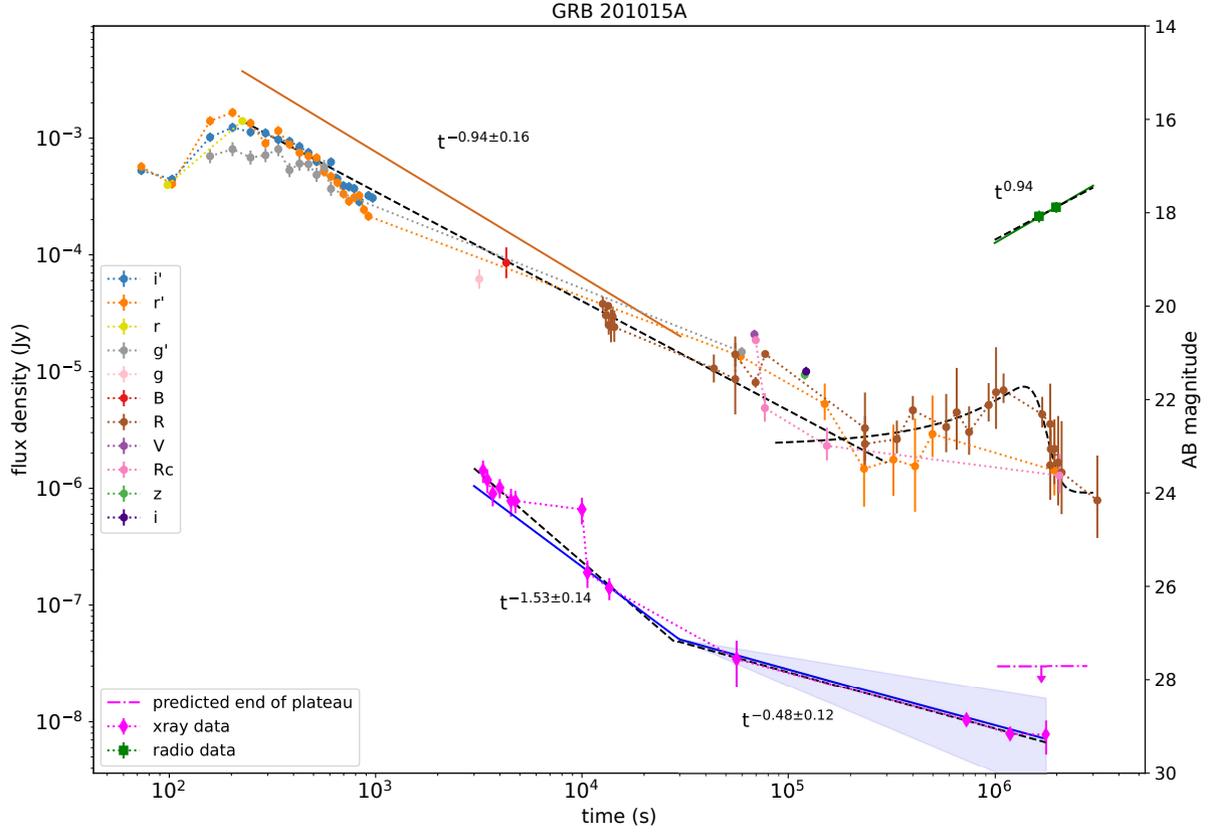}
\caption{A plot of the afterglow observational data (dotted lines) of the GRB 201015A with the best fitting power-laws and supernova model (black dashed lines) and Closure Relation (solid coloured lines) \citep{Gao13}. The optical afterglow of GRB 201015A is found from data published in GCNs, showing a mean power-law decay constant of $-0.94 \pm 0.16$ after the peak at $226 \pm 26$ s. The X-ray data from \emph{Swift} XRT and Chandra is fitted to a broken power-law with the decay constants $-1.53 \pm 0.14$ and $-0.48 \pm 0.12$ with a break at 26 ks. Radio data collected by e-MERLIN, also taken from GCNs, gives a decay constant of $\sim 0.94$. 
The closure relations are plotted with the mean value of the electron spectral index $p=2.42^{+0.44}_{-0.30}$, and the blue shaded region represents the upper and lower decay constants from the lower and upper confidence regions of the energy injection coefficient $q=0.24 ^{+0.24}_{-0.18}$. 
The pink arrow represents the prediction, based on the Dainotti Relation, for the end time of the plateau phase of the afterglow with the dash-dot line expanding to the 1-$\sigma$ lower and upper confidence region  \citep{Dainotti15}.}
\label{fig:Afterglow_lc}
\end{center}
\end{figure*}

We use the temporal and spectral indices described in Section~\ref{sec:obs}, which are summarised below and in Table~\ref{tab:CR}, to compare the observational data to the theoretical framework. The X-ray light curve is best described by a broken power-law with $\alpha_{\rm x1} = 1.53 \pm 0.14$ and $\alpha_{\rm x2} = 0.48 \pm 0.12$. The break indicating the start of the plateau phase is at $t_{\rm b} = (2.61 \pm 1.27) \times 10^4$~s. The spectral index of the first phase is $\beta_{\rm x1} = 1.16 \pm 0.22$, and $\beta_{\rm x2} = 1.10 \pm 0.13$ for the second phase. The optical light curve after the peak at $t_{\rm d} = 226 \pm 26$~s has a decay of $\alpha_{\rm o} = 0.94 \pm 0.16$ and the spectral index is measured to be $\beta_{\rm o} = 0.75 \pm 0.38$ from $\sim 150 - 600$ s. We could not get spectral information after the $t_{\rm b}$ time because there were no simultaneous observations with different filters as provided by the NUTTelA-TAO BSTI for the early time data. The temporal index after the break could not be determined due to the supernova. The radio provides late time information in a different waveband after $t_{\rm b}$ demonstrating a temporal index of $\alpha_{\rm r} \approx -0.94$. The light curve is shown in Figure~\ref{fig:Afterglow_lc}. The closure relations (CR) can provide a diagnostic for the surprising late-time rebrightening in the X-ray.

To find the electron spectral index $p$, we calculated it independently using the $\alpha$ and $\beta$ values for the optical and X-ray observations before the break time $t_{\rm b}$, and found the average for the scenario where the results are consistent with each other. We considered the wind to ISM transition as a possible explanation for the steep-to-shallow transition such as the case with GRB 140423A \citep{Li20}. However this failed to converge in consistent values of $p$. The best case scenario was relativistic, isotropic, self-similar deceleration phase for $\nu_a < \nu_m < \nu_c$ in ISM, in the slow cooling regime, with $p = 2.42 ^{+0.44}_{-0.30}$. 

To explain the afterglow after the break, we needed to introduce energy injection.
The shallowing of the light curve can be explained by the process of continuous energy injection from the central engine, possibly an accreting supra-massive neutron star, or magnetar or black hole \citep{Chen17, Li18}. 
The luminosity of this central engine is given by \citep{Zhang01}
\begin{equation}
    L = L_{0} \left(\frac{t}{t_0}\right)^{-q}
\end{equation}
where $t_{0}$ is the time at which the self-similar solution forms and the external shocks begin to decelerate, and $q$ is the energy injection parameter \citep{Zhang01,Li18}. Evidence of the injection would be present in the afterglow when $q<1$. Values of $q \sim 0.3$ are typical for GRBs expected to have energy injection from a magnetar or black hole \citep{Li18}. 
Another possibility is the energy injection from slower shells of ejecta catching up to the initial decelerating shock wave \citep{Zhang06}. In this scenario, the energy injection is described as $E_{\rm iso} \propto \gamma^{1-s}$ where $\gamma$ is the Lorentz factor, and $s$ is the shell model energy injection parameter.
We calculated the value of $q$ from the energy injection equation in \citet{Gao13}:
\begin{equation}
    \alpha = \frac{(2p-4)+(p+2)q}{4}
\end{equation}
using the measured value of $\alpha_{x,2}$ and the value of $p$ found above. The shallowing of the light-curve in the X-ray can be explained by energy injection parameter $q = 0.24 ^{+0.24}_{-0.18}$. The 1-sigma confidence range of $q$ relates to $\alpha_{x,2} = 0.48 ^{+0.26}_{-0.20}$. This range is denoted by the blue shaded region in Figure~\ref{fig:Afterglow_lc} along with the slopes for each of the light curves from the CR. The spectral and temporal indices given by the described CR are given in Table~\ref{tab:CR}

\begin{table*}
    \centering
    \begin{tabular}{|c|c|c|c|c|c|}%\cline{3-6}
    \hline \hline
    \multicolumn{2}{|c|}{{}}&\multicolumn{2}{|c|}{Temporal index $\alpha$} &\multicolumn{2}{|c|}{Spectral index $\beta$} \\\hline
    Temporal phase & Frequency & CR model & Observed & CR model & Observed \\\hline
    \multirow{2}{3cm}{\centering 
    No energy injection \\($<\sim3\times10^4$ s)} & 
    X-ray ($\nu > \nu_{c}$)&
    1.32 &$1.53\pm0.14$ &
    1.21 & $1.16 _{-0.23}^{+0.24}$ \\\cline{2-6} &
    optical ($\nu_m < \nu < \nu_c$) & 
    1.07 & $0.94\pm0.16$ 
    &0.71 &$0.75 \pm 0.39$ \\\cline{1-6}
    \multirow{3}{3cm}{\centering 
    Energy injection $q=0.3$ \\($>\sim3\times10^4$ s)} & X-ray ($\nu > \nu_{c}$)& 
    0.48 & $0.48\pm0.12$ & 
    1.21 & $1.10 \pm 0.13$ \\\cline{2-6} & 
    optical ($\nu_m < \nu < \nu_c$)&
    0.04 & - &
    0.71 & -  \\\cline{2-6}& 
    radio ($\nu_a < \nu < \nu_m$) &
    -1.13 & $\sim -0.94$ &
    -0.33 & - \\\hline \hline
    \end{tabular}
    \caption{The spectral and temporal indices in the best-fitting closure relation for this afterglow (relativistic, isotropic, self-similar deceleration phase for $\nu_a < \nu_m < \nu_c$ in ISM, with $p=2.42^{+0.44}_{-0.30}$), compared with the measured values for the relevant frequency ranges, both before and after energy injection at $\sim3\times10^4$s with $q=0.24 ^{+0.24}_{-0.18}$.}
    \label{tab:CR}
\end{table*}

\subsection{Supernova}\label{sec:SN}
%SN analysis
\begin{table*}
    \centering
    \begin{tabular}{|c|c|c|c|c|c|}
    \hline \hline
         SN model parameters & $t_0$ (s) & $\tau_{\rm rise}$ (s) & $\tau_{\rm fall}$ (s) & $A$ & $c$ \\ \hline
         fitting results & $(1.596 \pm 0.088) \times 10^{6}$ & $(1.17 \pm 0.35) \times 10^{5}$ & $(1.37 \pm 0.49) \times 10^{5}$ & $(9.8 \pm 2.4) \times 10^{-6}$ & $(9.2 \pm 3.4) \times 10^{-7}$ 
    \\\hline \hline
    \end{tabular}
    \caption{Resulting parameters from the SN model (Equation \ref{eq:SN}) fitting to the optical data beyond 1 day ($t > 86400$s)}
    \label{tab:SN}
\end{table*}

The optical observations later than 1 day show evidence of a supernova. The following analytic supernova model used in \citet{Taddia15} and \citet{Bazin11} was fit to the data:
\begin{equation}
    F(t) = A\frac{e^{\left(-\frac{t-t_0}{\tau_{\rm fall}}\right)}}{1+e^{\left(-\frac{t-t_0}{\tau_{\rm rise}}\right)}}+c
\label{eq:SN}
\end{equation}
where $F(t)$ is the flux density, $t$ is the time, $A$ and $c$ are normalisation constants, and $t_0$, $\tau_{\rm fall}$ and $\tau_{\rm rise}$ are related to the time of the peak of the supernova:
\begin{equation}
    t_{\rm peak} = t_0 + \tau_{\rm rise} \ln{ \left( \frac{\tau_{\rm fall}}{\tau_{\rm rise}}-1 \right)}
    \label{eq:SN_peak}
\end{equation}
The result of the fitting is given in Table~\ref{tab:SN} with 1 sigma errors, and presented in the light curve in Figure~\ref{fig:Afterglow_lc}. Based on the $t_0$, $\tau_{\rm fall}$ and $\tau_{\rm rise}$ values, the peak of the supernova is at $t_{\rm peak} = (1.4 \pm 0.4) \times 10^{6}$ s = $16 \pm 5$ days. The definition for the rise time given in \cite{Taddia15} is $t_{\rm rise} = t_{\rm peak} - t_{\rm expl}$ where $t_{\rm expl}$ is the average between the last non-detection of the supernova and the first detection point. Since we have observed the GRB and all times are given relative to the trigger time, by definition $t_{\rm expl} = 0$, and $t_{\rm rise} = t_{\rm peak}$.  The observed rise time of the SN converted to the rest frame of the GRB is $t_{\rm rise} = 11.3 \pm 3.5$ days. Based on the results from \citet{Taddia15}, this is within the 50$^{\rm th}$ percentile for SNe Ic and within the 11$^{\rm th}$ percentile for SNe Ic-BL. \citet{Taddia15} shows that SNe Ib or IIb have rise times $15 \lesssim t_{\rm rise} \lesssim 30$, making this SN more consistent with type Ic and Ic-BL. 

%\textbf{
The peak luminosity of the SN from our fits was $(7 \pm 5) \times 10^{-6}$ Jy, or $21.7 \pm 0.7$ mag, assuming a luminosity distance $D_L = 2520.1$ Mpc, we have an absolute magnitude of $M_{R} = -20.3 \pm 0.7$ mag. \citet{Lyman14} show that the bolometric correction for Type I core-collapse supernovae to $V$--band (approximate rest-frame of our observed $R$--band) around peak is close to zero (typically from $-0.2$ to $0.7$ mag), and so our $R$--band peak luminosity can be approximated to bolometric. Peak bolometric magnitudes range from approximately $-16$ to $-19$ mag for Type Ic-BL SNe \citep{Lyman16, Prentice19}. Although this indicates the SN associated with GRB 201015A was somewhat luminous in comparison, other luminous examples of GRB-SNe exist such as SN2011kl \citep{Kann16} and SN2012bz \citep{Schulze14a} and are more comparable with peak luminosities $M_V = -19.28$ mag and $M_V = −19.7$ mag respectively. 
SN2011kl is spectrally more similar to a super-luminous SN (SLSN) than other GRB-SNe, and it is associated with a ultra-long GRB (111209A) \citet{Greiner15}. Whereas SN2012bz is the counterpart of GRB 120422A, a relatively low luminosity GRB with a $T_{\rm 90} = 5.4 \pm 1.4$~s and \Eiso{}$= (1.6--3.2) \times 10^{50}$ erg \citet{Schulze14a}, and hence quite comparable to GRB 201015A.
Given our large statistical uncertainty on the peak, and an additional bolometric correction systematic uncertainty, we cannot analyse the luminosity of the SN associated with GRB 201015A beyond these general statements.
%}

\subsection{Dainotti Relation}
The canonical X-ray afterglow of GRBs contains a plateau phase where the decay constant increases, which is typically linked to energy injection \citep{Nousek06}. In the afterglow of GRB 201015A, the energy injection phase is seen to last surprisingly long and there is no visible end time. The 3D Dainotti relation links the end time of the plateau phase in the afterglow ($T_{\rm a}$), with the end of plateau X-ray luminosity ($L_{\rm a}$), and with the luminosity of the brightest second of the GRB prompt emission ($L_{\rm p}$) \citep{Dainotti17}. Since we do not know where the end of the plateau phase is, the separate 2D relations given in \citet{Dainotti15} were used in order to get a prediction of $T_{\rm a}$ and test if this afterglow violates the Dainotti relation. Combining the following equations from \citet{Dainotti15}:
\begin{equation}
    \log(L_{\rm a})=A+B\log (L_{\rm p})
\end{equation}
\begin{equation}
    \log(L_{\rm a})=\log(a)+b\log(T_{\rm a})
\end{equation}
gives: 
\begin{equation}
    \log(T_{\rm a}) = \frac{A + B\log(L_{\rm p})-\log(a)}{b}
    \label{eq:Ta}
\end{equation}
where $A = -14.67 \pm 3.46$, $B=1.21^{+0.14}_{-0.13}$, $b=-0.90^{+0.19}_{-0.17}$, $\log(a)=51.14\pm0.58$.
The peak luminosity, $L_{\rm p} = (5.81^{+1.68}_{-1.53})\times 10^{49}$ erg s$^{-1}$, was found from the peak 1 second spectrum of the prompt emission of GRB 201015A, created using the automated HEASoft {\sc batgrbproduct} processing. Using Equation~\ref{eq:Ta}, we find $T_a = 1.67 ^{+1.14}_{-0.65}\times 10^6$ s. This result is represented in Figure~\ref{fig:Afterglow_lc} by the position of the pink arrow, and the 1 sigma confidence region is indicated by the dash-dotted pink horizontal line.

The position of the predicted end time of the plateau falls closely with the last XRT observation. This long-lasting, and late shallow plateau phase could be a normal feature of a burst with a low \Eiso{} assuming these bursts follow the Dainotti relation, however we have not observed many late plateaux. 
%The observational bias against detecting low energy bursts, could have led to an observational bias against bursts with late plateau end times (beyond $10^4$ s) because of this correlation. Another challenge which could be skewing our statistics on the typical end times of plateaux is that later plateaux would be intrinsically dimmer because of the decay of the afterglow.

\subsection{Jet opening angle}
The observation of a jet break - a shallow-to-steep transition in the light curve - is a useful feature which the majority of GRB observations do not have \citep{Racusin09b}.  Based on the estimated end time for the X-ray plateau, we estimate the lower limit for the jet half opening angle, $\theta_j$. Here we assume the jet break time is $t_j \ge 1.76 \times 10^6$ s $= 20.4$ days, as this is the time of the last measurement in the X-ray band. The start time of the energy injection is estimated as $t_i = (2.61 \pm 1.27) \times 10^4$ s $= 0.3$ days based on the broken power-law fit to the X-ray light curve. Since we are finding the lower limit for $\theta_j$, we approximate the circumburst density to be $n_0 = 0.1$ taking the lower value from the range of $n_0 = 0.1$ -- $100$ cm$^{-3}$ for long GRBs given in \citet{Panaitescu02}. The energy injection parameter has been converted from the q-value to the s-value for the shell model using the conversion equation for the ISM shock model in \citet{Zhang06}:
\begin{equation}
    s = \frac{10-7q}{2+q}
\end{equation}
giving $s=3.7$ for $q=0.24$. 
The initial isotropic kinetic energy of the burst is given by:
\begin{equation}
    E_{\rm K,iso} = E_{\gamma,\rm iso}\left(\frac{1}{\eta_\gamma}-1\right)
\end{equation}
hence $E_{\rm K} \sim 0.158 \times 10^{52} $ erg if we take $\eta_\gamma$ to be approximately $0.1$. 
To calculate the total kinetic energy we use the following equation:
\begin{equation}
    E_{\rm K,final} = E_{\rm K,intial} \left(\frac{t_f}{t_i}\right)^{3(s-1)/(7+s)}
\end{equation}
providing the final isotropic kinetic energy $E_{\rm K} \sim 3.8 \times 10^{52}$ erg.
Finally, using the equation for half jet opening angle in \citet{Fong14}:
\begin{equation}
    \theta_j = 9.51 t_{\rm j,d}^{3/8} (1+z)^{-3/8} E_{\rm K,iso,52}^{-1/8} n_0^{1/8} \rm deg 
\end{equation}
shows that $\theta_j \ge 16^{\circ}$. This value of $\theta_j$ is consistent with the population of long GRBs which have a range of $2^{\circ} < \theta_j < 25^{\circ}$, although it is greater than the mean of $\theta_j = 7^{\circ}$ \citep{Fong14}. We have treated this as a uniform jet, but other jet structures and components (e.g. cocoon) may provide different results \citet{Lamb21}. 

Assuming the half jet opening angle, $\theta_{j} = 16^{\circ}$, we can estimate the energy emitted in gamma-rays from the prompt emission of this GRB. The \Eiso{} is corrected by the beaming factor \citep{Peng05}:
\begin{equation}
    E_{\gamma} = E_{\gamma,\rm iso} \left( \frac{\theta_{j}^{2}}{2} \right)
\end{equation}
giving $E_{\gamma} \sim 6.7 ^{+2.3}_{-2.0} \times 10^{48}$~erg.

\section{Comparison to the spectrally soft population}\label{sec:softpop}

Only three GRBs in the third \emph{Swift}-BAT catalogue \citep{Lien16} have a prompt emission spectrum with photon indices $\Gamma > 3$: 050416A, 080520 and 140622A.
From this sample, 140622A is the only GRB thought to have a merger progenitor, based on the duration $T_{\rm 90} = 0.13 \pm 0.04$ s, although this characterisation as a short GRB is uncertain based on its unusually soft spectrum \citep{GCN16438}. Machine learning analysis of the prompt emission also categorises it as a short GRB \citep{Jespersen20}. The others have been classified as long GRBs based on their $T_{\rm 90}$ being greater than 2 seconds and having a soft spectrum, however without the observation of a supernova or kilonova counterpart, these classifications cannot be confirmed.

%The morphology of the 140622 burst differs from the other bursts in this sample when comparing the light curves at different energies shown in %Figure~\ref{fig:promptLC}. The plot was created by {\sc batbinevt} with 64ms binning for the four energy bands: 15 -- 25 keV, 25 -- 50 keV, 50 -- 100 keV, 100 -- 150 keV. GRB 140622A is the only burst to show significant emission in 50 -- 100 keV with a short hard spike, whereas the other bursts exhibit longer activity with multiple flares in the lower energy bands. The bursts are all relatively weak in the 100 - 150 keV band, and GRB 080520 appears weak in all energy bands, possibly due to its higher redshift of $z=1.55$ [GCN 7757] while the rest of the bursts have a redshift of $z < 1$. \\
%\rlcs{How confident are you that 140622 has significant 50-100 keV emission and the others don't? It isn't very obvious to me from Fig 3 and I also wonder if there is a better way to represent the light curve behaviour to make your point because with all 4 overlaid it's hard to see the differences. If we keep it, can you add the energy bands to the y-axis labels and keep the same y-height but make it full page width? would that help for making out detail? Or have one panel of just 201015A to show the spike+EE and represent the energy bands and the others some other way. sorry not sure!}

We fit the BAT $T_{\rm 100}$ spectra from the third \emph{Swift}-BAT catalogue \citep{Lien16}, and XRT time-averaged spectra from the \emph{Swift} Burst Analyser \footnote{\url{https://www.swift.ac.uk/burst_analyser/}} \citep{Evans10} to a simple power law model to find the photon indices. We find that the photon indices for both the BAT and XRT bands are similar for each of the GRBs: $\Gamma_{\rm BAT} \sim 3$ and $\Gamma_{\rm XRT} \sim 2$. The values of each of the photon indices can be found on Table~\ref{tab:EpEiso}.

\begin{figure*}
\includegraphics[width=\textwidth]{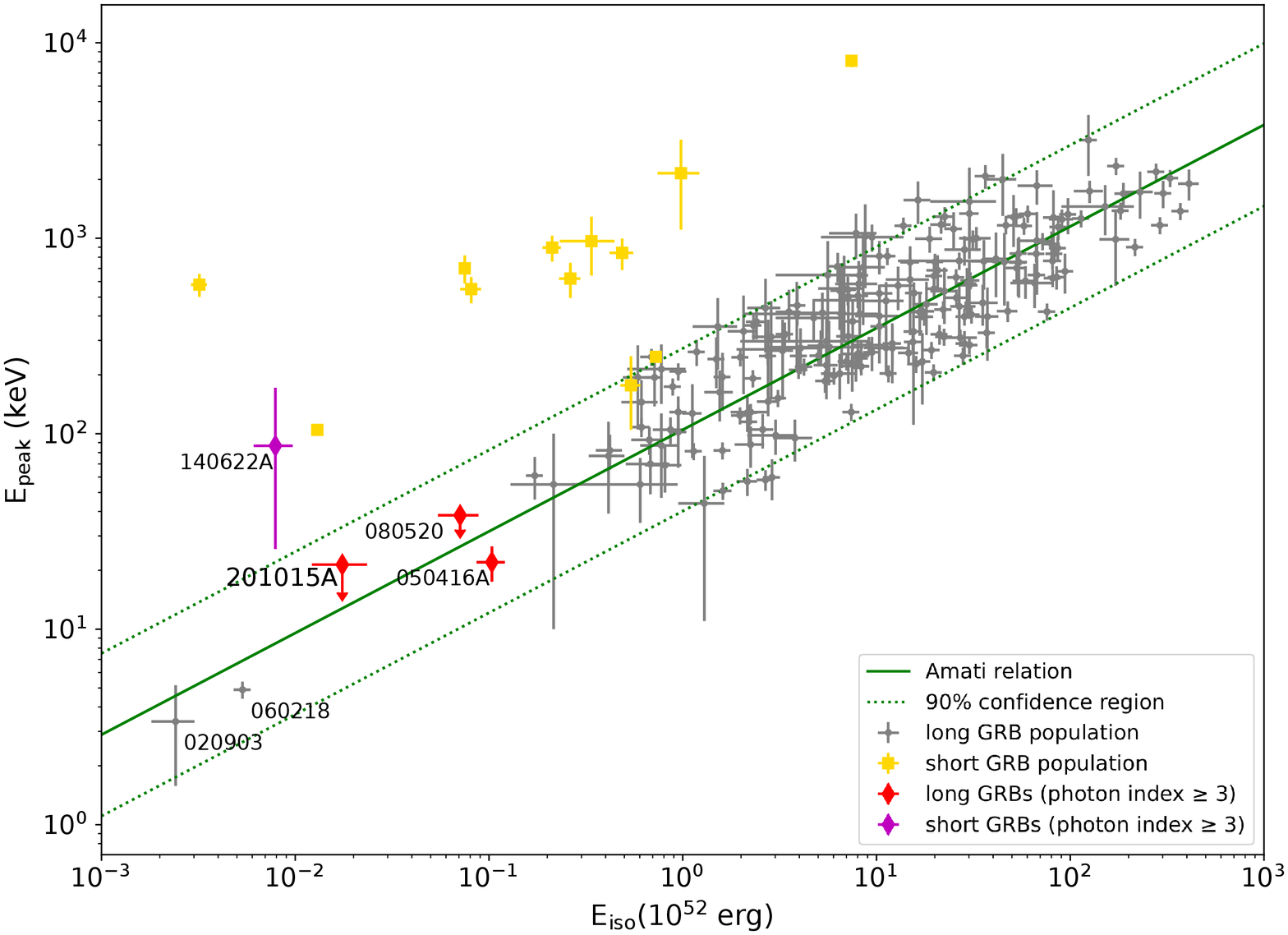}
\caption{The equivalent isotropic energy \Eiso{} ($10^{52}$ ergs) of GRBs plotted against their spectral peak energy $E_{\rm p}$ (keV), showing the correlation known as the Amati relation (green lines) \citep{Amati06}. The grey points represent the general population of long GRBs detected by various instruments, and the yellow squares represent the short \emph{Swift} GRBs presented in \citet{D'Avanzo14}. The spectrally soft sample of GRBs  with photon index ($\Gamma) \ge 3$ is marked in red for long GRBs and purple for short GRBs.}
\label{fig:Amati}
\end{figure*}

The Amati relation provides a correlation between $E_{\rm p}$ and \Eiso{} of the time-averaged prompt spectrum of long GRBs \citep{Amati02}. 
In order to check the $E_{\rm p}$ and \Eiso{} of this sample of GRBs relative to the Amati relation, we have selected the best fitting model from a simple power-law, cut-off power-law, and Band function fit to the $T_{\rm 100}$ BAT spectra. The method used is as described in Section~\ref{section:gamma} for GRB 201015A, where the equation for the spectral peak energy (Equation \ref{eq:Ep}) uses the low-energy spectral index ($\beta_1$) in the case of the Band function. The values calculated for GRB 201015A and the other 3 GRBs in this sample were plotted onto an Amati plot (Figure~\ref{fig:Amati}) with a large sample of long GRBs with well measured redshift and spectral parameters \citep{Amati06}, as well as a few \emph{Swift} short GRBs for comparison \citep{D'Avanzo14}. We have an upper limit for $E_{\rm p}$ in the case of GRB 201015A and GRB 080520 because the cut-off power-law or Band function could not be adequately fit to their spectra using {\sc xspec}. The lower-limit for the bandpass of the BAT detector (15 keV) was considered the upper limit for the break, and then converted to the rest frame of the GRB using the redshifts of the GRBs.
Figure~\ref{fig:Amati} clearly indicates that these bursts along with 2 others (060218, and 020903) are outliers in terms of their low-\Eiso{} and low-$E_{\rm p}$ compared to the rest of the population, but they still fit the Amati relation. The findings for these low-$E_{\rm p}$ and low-\Eiso{} events are shown in Table~\ref{tab:EpEiso}. We find that this population is of low-redshift with 5 out of 6 GRBs having a redshift of $z < 1$. Most of the GRBs have short timescales $T_{\rm 90} < 10$ s, apart from the GRB 060218 $T_{\rm 90}$ calculated by \citet{Campana06}.

%\bg{Maybe we could quantify how many sigma from the sample mean this is?}.\\

\begin{table*}
\centering
\begin{tabular}{|c|c|c|c|c|c|c|}
\hline\hline
GRB & Redshift & $T_{\rm 90}$ (s) & \Eiso{} ($\times10^{50}$ erg) & $E_{\rm p}$ (keV) & BAT photon index & XRT spectra photon index\\
\hline
201015A & 0.426 & $9.78 \pm 3.47$ & $1.75 ^{+0.60}_{-0.53}$& $<21.39$ & $3.00^{+0.50}_{-0.42}$& $2.16^{+0.41}_{-0.36}$\\[1mm]
%\hline
140622A & 0.959 & $0.13 \pm 0.04$ & $0.79 \pm 0.18$ & $87.3 ^{+77}_{-63}$& $3.17 ^{+0.20}_{-0.17}$& $1.70 \pm 0.50$\\[1mm]
%\hline
080520 & 1.55 & $3.32 \pm 0.86$ & $7.10^{+1.73}_{-1.65}$ & $< 38.25$& $3.14^{+0.42}_{-0.35}$& $2.10^{+0.33}_{-0.31}$\\[1mm]
%\hline
060218 & 0.033 & $2100 \pm 100$\footnote{\citep{Campana06}} & $0.534 \pm 0.053$ & $4.9 \pm 0.49$& $2.18 ^{+0.20}_{-0.18}$ & $1.67 \pm 0.01$\\[1mm]
%\hline
050416A & 0.6535 & $6.7 \pm 3.4$ & $10.3^{+1.7}_{-1.7}$ & $22.0^{+4.5}_{-4.5}$& $3.27 ^{+0.21}_{-0.19}$& $1.96^{+0.10}_{-0.09}$\\[1mm]
%\hline
020903 & 0.25 & $\sim 3.3$ & $0.24 \pm 0.06$ & $3.37 \pm 1.79$&-&-\\
\hline\hline
\end{tabular}
\caption{Tabulated values for the bursts of interest in the low-$E_{\rm peak}$ and low-\Eiso{} section of the Amati plot. The $T_{\rm 90}$ values have been gathered from GCN notices, except 060218, which was from \citet{Campana06}. We calculated the \Eiso{} and $E_{\rm peak}$ using {\sc xspec} unless already provided in \citet{Amati06}. The BAT and XRT photon indices were also calculated using {\sc xspec}, except for 060218, which was gathered from the third \emph{Swift}-BAT catalogue \citep{Lien16} and the \emph{Swift} Burst Analyser \citep{Evans10}.}
\label{tab:EpEiso}
\end{table*}

\begin{figure*}
    \centering
    \includegraphics[width=15cm]{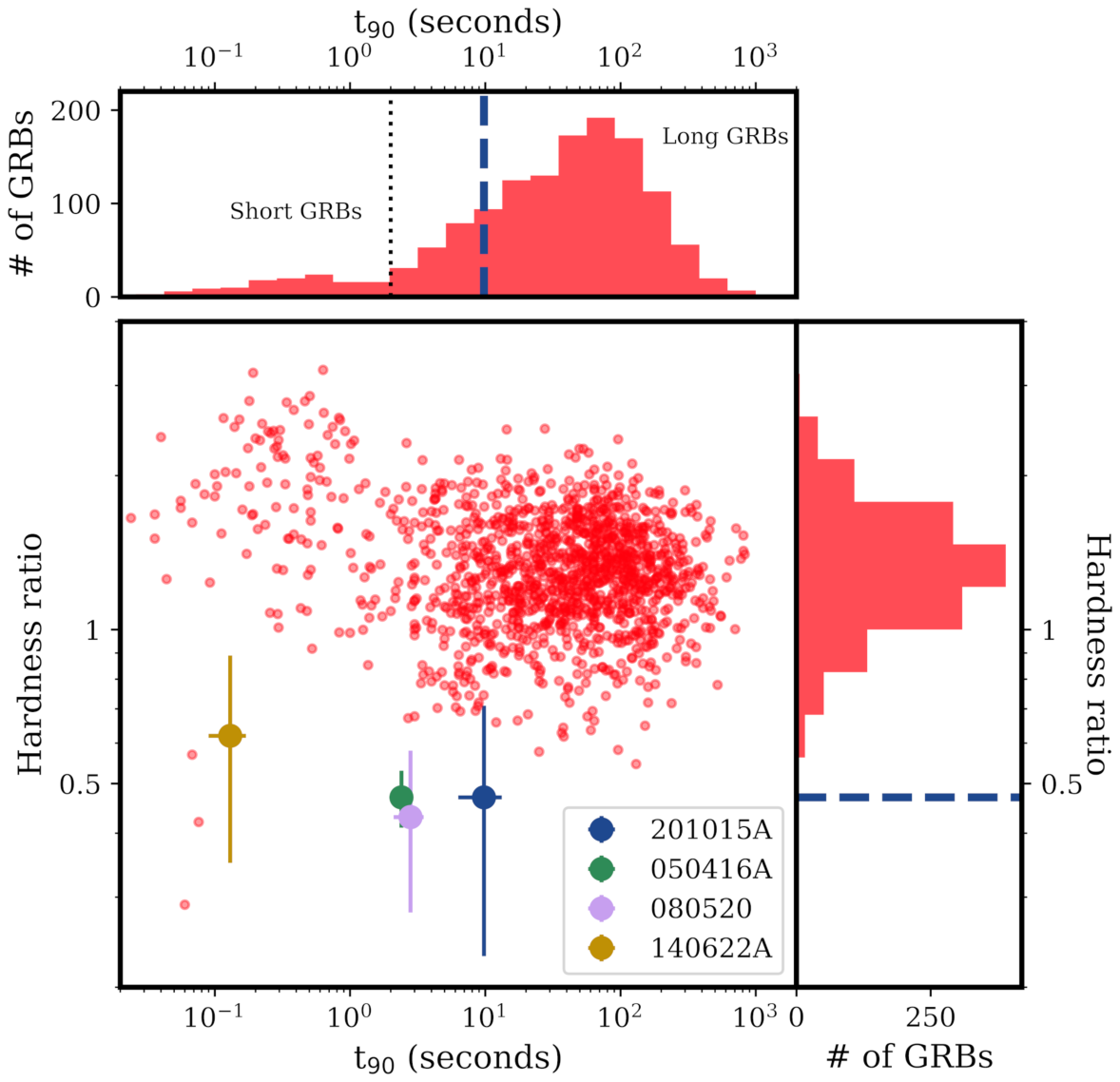}
    \caption{Plot of $T_{\rm 90}$ against the hardness ratio (fluence in 50 -- 100 keV band over fluence in 25 -- 50 keV band from the power-law fit) of GRBs from the third \emph{Swift}-BAT catalogue \citep{Lien16}. The GRBs with photon index $\Gamma > 3$ are coloured differently on the plot and labelled in the legend. The histograms of $T_{\rm 90}$ and HR are shown on the top and right sides of the plot, with the properties of GRB 201015A highlighted with a dashed blue line. }.
    \label{fig:T90_HR}
\end{figure*}

We have used the data from the \emph{Swift} BAT catalogue \citep{Lien16} to plot these GRBs on a plot of $T_{\rm 90}$ against hardness ratio (HR). To find HR, we have used the simple power law fit for all the GRBs in the sample to calculate the ratio of the fluence in the 50 -- 100 keV band to the 25 -- 50 keV band. For GRB 190326A, there was no error provided for the fluence or the fit parameters. Figure~\ref{fig:T90_HR} shows the scatter plot of the $T_{\rm 90}$ and HR values, with histograms of both distributions. By fitting a gaussian curve to the HR distribution, we find that the HR for GRB 201015A is 3.4 sigma from the mean and the GRB 140622A HR of 0.62 is 2.5 sigma from the mean. In Table~\ref{tab:T90_HR}, we have tabulated a list of soft GRBs with HR $\leq 0.62$ and $T_{\rm 90} < 10$ s, with information on the classification of the GRBs and their redshift found from GCN circulars. The sample of soft GRBs are a mix of mergers and collapsars, with shorter durations than the the general population of short and long GRBs respectively. These GRBs are also found at lower redshifts than most detected GRBs, since only 10\% of GRBs have $z<1$ \citep{Le17}.

\begin{table}
    \centering
    \begin{tabular}{|c|c|c|c|c|}
         \hline \hline
         GRB &$T_{\rm 90}$ (s)&HR&classification & redshift\\
         \hline
050416A	&$2.49 \pm 0.44$&$0.44 \pm 0.06$& long&0.6535\\
080520	&$2.82 \pm 0.66$&$0.43 \pm 0.15$& long&1.545\\
090417A	&$0.068 \pm 0.022$  &$0.57_{-0.16}^{+ 0.15}$& short&0.088\\
140622A	&$0.13 \pm 0.04$&$0.62 \pm 0.27$& short&0.959\\
150101A	&$0.060 \pm 0.011$  &$0.29\pm0.16$& short&-\\
%150101B	&0.012	&0.039& short&0.134\\
180718A	&$0.084 \pm 0.023$	&$0.16\pm0.09$& short&-\\
190326A	&$0.076 \pm 0.032$	&$0.42\pm $N/A& short&-\\
201015A &$9.78 \pm 3.47$&$0.47\pm0.15$ & long & 0.426 \\
\hline \hline
    \end{tabular}
    \caption{List of soft GRBs in the BAT catalogue with a low hardness ratio ($<0.6$) and $T_{\rm 90}$ less than 10s \citep{Lien16}, with redshift information collected from GCN notices.}.
    \label{tab:T90_HR}
\end{table}

\begin{figure*}
\includegraphics[width=\textwidth]{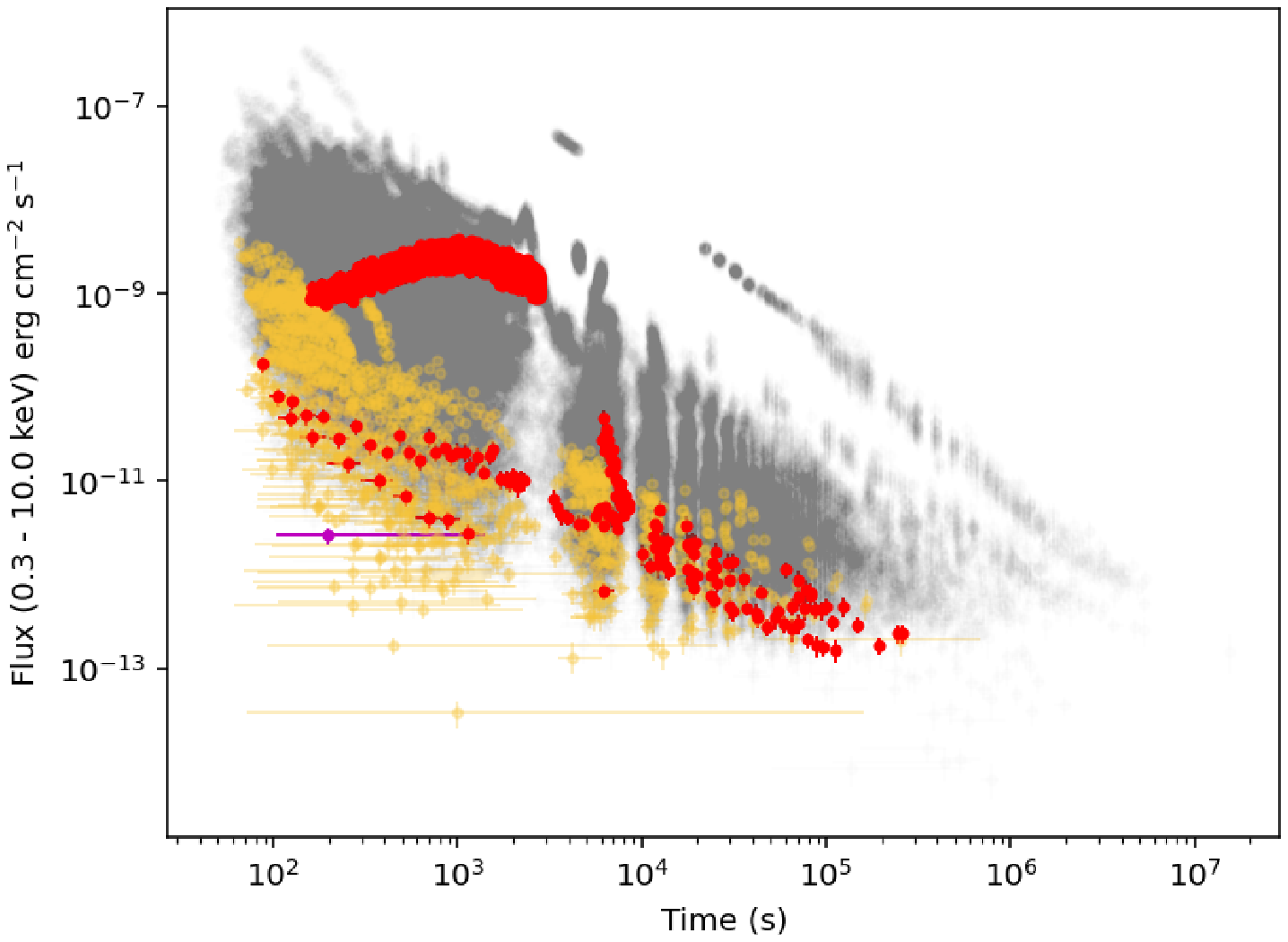}
\caption{%\textbf{
The XRT afterglows of Swift GRBs shown here are presented in flux units. This includes the light curves of GRBs 060218, 080520, 050416A, and
201015A in red, and GRB 140822A in purple, overlaid on GRBs with $T_{\rm 90}$ > 2s in grey, and GRBs with $T_{\rm 90}$ < 2s in yellow.
}
%}
\label{fig:XRT_afterglows}
\end{figure*}

%\textbf{
The sample of low-luminosity soft GRBs detected by \emph{Swift} do not seem to show any similarity in their morphology in the prompt emission light curves. To compare their X-ray afterglows, we have plotted their afterglow light curves in Figure~\ref{fig:XRT_afterglows} showing the flux in the 0.3--10.0 keV band, along with all \emph{Swift} GRBs followed up by XRT. This plot was created using the {\sc swifttools} API\footnote{\url{https://www.swift.ac.uk/API}} \citep{Evans07}.  The median light curve of short GRBs (yellow) is lower than that of long GRBs (grey) as shown in \citet{Margutti13}.  This figure shows that the low-$E_{\rm p}$ and low-\Eiso{} long GRBs (red) typically have X-ray afterglows with lower flux than the median of the population of long GRBs, and are more in line with the short GRB median light curve. The short GRB 140822A (purple) only has one data point, which is also lower in flux than the median for short GRBs, however this is not enough data to draw conclusions. The outlier in the low-$E_{\rm p}$ and low-\Eiso{} long GRBs is 060218 which is more luminous in the 0.3--10.0 keV range than the other GRBs in red up to $\sim 10^{4}$~s after which it joins the others. This emission has been attributed to the prompt emission of the GRB rather than the afterglow since it has $T_{\rm 90} = 2100\pm100$ s \citep{Campana06}.%}

GRB 060218 has a remarkably low redshift of $z = 0.033$, but is the outlier in terms of the $T_{\rm 90}$. The duration of the burst is unclear based on the {\sc batgrbproduct} automated analysis\footnote{\url{https://gcn.gsfc.nasa.gov/notices_s/191157/BA}}, but \citet{Campana06} has determined $T_{\rm 90} = 2100 \pm 100$~s for this burst. It is believed to be off-axis by a few degrees based on the inner fast variability being attributed to a precessing jet \citep{GCN4819}. This GRB was associated with supernova SN 2006aj based on the continuum spectrum of the optical transient detected by VLT \citep{GCN4803}. The underluminous nature of this burst and the detection of a thermal component in the X-ray is interpreted as a mildly relativistic jet causing a shock breakout into dense circumstellar material \citep{Campana06, Nakar15}. An alternative model suggests a low-luminosity, low Lorentz factor jet ($\gamma = 10$) with a low-mass circumstellar envelope, and extinction due to dust \citep{Irwin16}. GRB 100316D is an analogous burst to GRB 060218 with a low $E_{\rm peak}$ in the range of 10 -- 42 keV (90\% confidence range) and an estimated \Eiso{}$ \ge (5.9 \pm 0.5 ) \times 10^{49}$~erg \citep{Starling11}. Unlike these GRBs, the spectrum of GRB 201015A is not well fitted with a blackbody, making it unlikely to have a thermal component.

XRF 020903, a transient detected in the 2 -- 5 keV band by the WXM instrument on HETE observing in the 2 -- 30 keV range \citep{GCN1530}, was categorised as an X-ray Flash (XRF) due to its extremely low $E_{\rm p}$ of $3.37 \pm 1.79$ \citep{Soderberg04}.  The event coincided with a SN confirmed by the rebrightening in the optical at 24 days and spectroscopic follow up of the optical transient \citep{GCN1554}. \citet{Urata15} studied the afterglow light curve discovering the achromatic rebrightening hence the off-axis viewing. Therefore this event is considered to be an off-axis orphan GRB, for which the narrow collimated relativistic jet is not seen, but the wider afterglow signal can still be detected. This is expected to be observed for GRBs with $\theta_{\rm obs} > 20^{\circ}$ \citep{Urata15}. Furthermore, spectroscopic observations into the host galaxy of XRF 020903 with ALMA identified similar properties to GRB host galaxies, supporting the origin of this event to be the same as long GRBs \citep{Chen21}.

%\rlcs{in the 060218 discussion, I note that GRB 100316D appears to be an analogue to that burst, and is one which we studied in detail (Starling et al. 2010 or 2011). It would be useful to have a look at that one in comparison, and see whether it does/does not fit into your 'soft' category - it certainly is low luminosity, as I recall, and requires a thermal component at early times in X-rays (but there are some thermal X-ray GRBs that are not LLGRBs - there is some literature on that which we don't want to go into unless there is evidence in 201015A for any thermal emission [could a hidden thermal component create the effect of an extended plateau?]).}

\section{Discussion}\label{sec:Discuss}

\subsection{Classification}
It is not immediately clear from the high energy prompt emission observations alone whether GRB 201015A belongs in the long or short GRB category. The BAT $T_{\rm 90} = 9.78 \pm 3.47$ s is consistent with a long burst, whereas the sub-threshold detection by \emph{Fermi} GBM provides 1.024 s as an estimate of the duration. This emphasises the uncertainty in the measured duration of the bursts based on the detector bandpass, and how this can lead to incorrect categorisation based on $T_{\rm 90}$. The \emph{Swift} BAT bandpass is 15--150 keV \citep{Gehrels04}, whereas \emph{Fermi} has a larger effective energy range of 8 keV--40 MeV \citep{Yu16}.

The upper limit of $E_{\rm p}$ and value of \Eiso{} for GRB 201015A are consistent with the `Amati relation' for LGRBs \citep{Amati02,Amati06,Minaev20}. However, the short pulse and extended tail morphology is more in line with a short or extended emission \citep[EE;][]{Norris06,Norris10} GRB. The measured \Eiso{} $= 1.75 ^{+0.60}_{-0.53} \times 10^{50}$~erg is under-luminous for an LGRB; the log mean \Eiso{} is $\sim 6.6 \times 10^{52}$~erg in the GBM-selected sample of \citet{Gompertz18b}. It is, however, more consistent with an SGRB. Mean \Eiso{} is $\sim 6.4 \times 10^{49}$~erg in the complete sample of SGRBs with redshift \citep{Gompertz20}. To check for consistency, we compared our independent fitting results against estimates from other sources such as GCNs, e.g. \citep{GCN28668} for the \Eiso{} estimate of $\sim 10^{50}$ erg for GRB 201015A.
\citet{D'Avanzo14} and \citet{Minaev20} show the similar correlation for SGRBs lies just above the Amati relation for LGRBs and some SGRBs may fall closer to the LGRB relation and vice versa so this relation is not a reliable method to determine the classification of a GRB. However, the 4 bursts plotted on Figure~\ref{fig:Amati} in colour do lie in their respective regions of the $E_{\rm p}$ and \Eiso{} correlations, showing this relation holds in the low energy range. Given the ambiguity in the distinction of $T_{\rm 90}$, HR, and \Eiso{} for LGRBs and SGRBs, future GRBs will be difficult to classify without a supernova or kilonova counterpart to confirm their origin.

\subsection{Low-luminosity soft bursts}
The relative difficulty of finding bursts with low luminosity, and low spectral peak energy and their afterglows results in bursts like these going undetected by instruments designed to trigger for GRBs with high keV energies. Off-axis events for which the prompt emission may not be visible but the afterglow signal could be, might not trigger detectors and therefore have no follow up to find the afterglow other wavebands. Future missions such as SVOM and Einstein Probe will be better equipped to detect bursts in this parameter space. Einstein Probe will launch in 2023 with the intention to discover energetic transients and variable objects in the X-ray band of 0.5 -- 4 keV with a large field of view: $60^{\circ} \times 60^{\circ}$ \citep{Yuan15}. SVOM is also set to launch in 2023, searching for low-energy bursts in the 4 -- 250 keV range using the Eclairs instrument with $89^{\circ} \times 89^{\circ}$ field of view, and the MXT instrument providing data in the 0.2 -- 10 keV energy band \citep{Bernardini21}. THESEUS is a potential future mission with the goal of finding high-redshift GRBs up to $z \sim 10$, possibly broadening our sample of soft GRBs at higher redshifts \citep{Amati21b}. Another proposed mission, Gamow Explorer, would trigger JWST and ground based telescopes for follow-up of LGRBs at $z>6$ for spectroscopic and multi-wavelength data \citep{White20}.

The population of GRBs exhibiting low $E_{\rm p}$ and low \Eiso{}, are found in a scarcely populated region of the Amati plot (Figure~\ref{fig:Amati}).
The low luminosity could be a result of lower intrinsic energy release from these bursts, or a choked jet due to circumburst material ejected from the supernova. Alternatively these values could be explained by an off-axis viewing angle such as with the famous GRB 170817A \citep{Lamb20} or more recently GRB 190829A \citep{Sato21}. XRF 020903 also shows evidence of off-axis viewing based on the achromatic rebrightening and is considered an off-axis orphan afterglow \citep{Urata15}. %\textbf{
The off-axis model should reduce the flux, but not affect the photon index of the GRB. It does however lower the characteristic frequency making the overall spectrum softer, and can make the observed photon index softer if the characteristic frequency is shifted across the observational energy band \citep{Lamb21b}. For GRB 201015A, we cannot find evidence of achromatic rebrightening so the jet is not considered to be off-axis.%}
In the case of GRB 060218, it is considered to be a result of a shock breakout into a dense circumburst environment by \citet{Campana06, Nakar15}, but interpretted as a lower Lorentz factor jet  by \citet{Irwin16} which makes the outflow more opaque to the gamma-rays. %\textbf{
The shock breakout model should not affect the photon index of the prompt emission of the GRB, so this provides an explanation of why this GRB has $\Gamma = 2.18^{+0.20}_{-0.18}$, a more typical value compared to $\Gamma \ge 3$ \citep{Irwin16}.%} 
GRB 060218 is thought to be off-axis by a few degrees, with the precession of the jet resulting in rebrightening and variation in the optical and radio afterglows observed \citep{GCN4819}.  For GRB 201015A there is no evidence of a shock breakout based on the spectral fitting, and it is difficult to determine whether there is rebrightening in the optical at the same time as the X-ray rebrightening due to the SN, indicating the most likely scenario is that of an intrinsically low luminosity jet. The relatively large half jet opening angle estimated to be $\theta_j >16^{\circ}$ suggested it is a poorly collimated jet. 
We find that there are various scenarios which make GRBs appear to have low-\Eiso{}, and for all these GRBs, follow-up observations are required to understand the nature of what is causing the lower luminosity.%\bg{Can we find some references for why this might be the case? Less collimated by the stellar envelope? Slower rotating progenitor star? Less massive star?}

\subsection{Follow up observations}
Based on the closure relations found to best fit the temporal and spectral indices measured for GRB 201015, we find that the energy injection scenario is the most plausible for explaining the X-ray light curve. GRB 130603B was also found to have energy injection characterised by $q=0.3$, but the radio data does not fit the energy injection model leading to the magnetar spin-down scenario being the best explanation \citep{Fong14}. Treating the shallowing of GRB 201015A as a normal GRB X-ray afterglow plateau, which is a phase linked to energy injection \citep{Bernardini12b}, shows that it does not violate the Dianotti relation even though it is a very late plateau \citep{Dainotti15}. It is difficult to explain the origin of this central engine activity and some relate this phase to slightly off-axis viewing \citep{Beniamini20}. The closure relation described in Section~\ref{section:CR} fits the afterglow very well, this is partially due to the energy injection coefficient chosen to match the X-ray data after the beginning of the plateau, but this also fits remarkably well with the radio observations, supporting the scenario of ongoing central activity at late-time from this GRB. The earlier phase of X-ray data seems to lie above the prediction from the closure relation which could be an indication of flares from central engine activity as commonly seen in some \emph{Swift} GRBs \citep{Zhang06}. 

%\textbf{
The TeV emission detected from GRBs is considered independent to the prompt emission as it is most likely produced by inverse Compton of the afterglow \citep{Zhang20c}, therefore we cannot explain the VHE emission with the models proposed for the low energy soft GRBs. More observations of GRBs in VHE are required for further study with comparisons of the afterglows, and prompt emission of VHE GRBs.%}

We have attempted to account for the variation in observations in the different optical bands using the zero points of the filters \citep{Frei1994}, after adjusting for galactic extinction and absorption based on assumed models for the wavelengths \citep{Schlafly11}, but it is difficult to account for host galaxy reddening because the extinction curve is unknown. This can also affect the measured optical spectral index we have used to infer the closure relations for this GRB.

Due to the ambiguous prompt emission, the follow-up observations in the optical showing evidence of a SN were necessary to identifying the classification of this burst. This GRB highlights the importance of follow-up missions in categorising GRBs and studying their jets and environments.

\section{Conclusions}\label{sec:Conclusion}

GRB 201015A is an interesting long GRB for many reasons. It is the fifth burst to have a candidate VHE detection, and is the burst with the lowest prompt energy out of this group. It has a surprisingly low-luminosity for a GRB with isotropic energy \Eiso{}$ = 1.75 ^{+0.60}_{-0.53} \times 10^{50}$ erg, and $E_{\gamma} \sim 6.7 ^{+2.3}_{-2.0} \times 10^{48}$ erg based on the predicted opening angle of the jet $\theta_j \ge 17^{\circ}$. Relating to the low \Eiso{}, the spectral peak energy is $E_{\rm p} < 21.39$ keV, and follows the Amati relation trend for long GRBs. The burst's prompt emission spectrum is unusually soft with a photon index of $\Gamma = 3.00 ^{+0.50}_{-0.42}$. The hardness ratio of this GRB is 0.47 which is 3.4 sigma lower than the mean for \emph{Swift} bursts.

We have compared GRB 201015A to others which are similarly spectrally soft ($\Gamma \ge 3$ and HR $\leq 0.62$) from the \emph{Swift} BAT catalogue. We find that the bursts with $\Gamma \ge 3$ are generally short-duration long GRBs which lie in the low-$E_{\rm peak}$, low-\Eiso{} region of the Amati plot. The sample with HR $ \leq 0.62$ introduces more short GRBs to our selection of soft GRBs. Both the short and long GRBs in this category tend to have a lower duration than the majority of the general short and long GRB populations, and the majority are of low redshift (all except GRB 080520 have $z<1$). Looking into the other GRBs in this parameter space shows that they have a range of explanations for their lower luminosity compared to the general population of GRBs. This includes off-axis viewing, shock-breakout into a denser envelope, or an intrinsically less energetic jet. The last scenario is what we have found to be the most likely for GRB 201015A due to lack of evidence for the other cases.

The afterglow of GRB 201015A was an important part of this study, since the detection of a SN in the optical bands confirmed the classification of this burst as a long GRB, and the X-ray data showed a peculiar steep-to-shallow transition at a late-time ($t_{b} = (2.61 \pm 1.27) \times 10^{4}$ s). The temporal and spectral indices found from the X-ray, optical and some radio observations were used to match with closure relations given in \citet{Gao13}. The best fitting closure relations were for the relativistic, isotropic, self-similar deceleration phase for the $\nu_a < \nu_m < \nu_c$ regime in the ISM (Table 13 in \citealt{Gao13}), and electron spectral index $p = 2.42 ^{+0.44}_{-0.30}$. We have also matched the plateau phase to energy injection with $q = 0.24^{+0.24}_{-0.18}$ after $t=(2.61\pm1.27)\times10^4$ s. The observed plateau phase of this X-ray afterglow was tested against the expected end time of the plateau $T_a$ from the Dainotti relations, and found to be $T_a = 1.67^{+1.14}_{-1.53} \times 10^6$ s which is around the last observation in the X-ray. This shows that this GRB does not violate the Dainotti relation \citep{Dainotti15}. We have then used the predicted $T_a$ to constrain the lower limit of the jet half opening angle $\theta_{j} \ge 16^{\circ}$. The SN associated with this GRB has a rise time $t_{\rm rise} = 11 \pm 1$ days, consistent with type Ic and Ic-BL \citep{Taddia15}.

With future missions such as SVOM and Einstein Probe searching in lower energy gamma-rays, and X-rays we will find more bursts like GRB 201015A with higher field-of-view and more sensitive instruments \citep{Yuan15, Bernardini21}. This will help us build a greater data set for statistical analysis of the soft GRB population. Follow-up observations of the afterglow in multiple wavelengths are also required to confirm classification of the GRB, and determine properties of the jet and environment.

\section*{Acknowledgements}

M.P., P.O., and G.P.L acknowledge the support by the UKRI Science and Technology Facilities Council (STFC). 
B.G. and M.N. are supported by the European Research Council (ERC) under the European Union’s Horizon 2020 research and innovation programme (grant agreement No.~948381). M.N. acknowledges a fellowship from the Alan Turing Institute.
RPB acknowledges support from the ERC under 550 the European Union’s Horizon 2020 research and innovation programme (grant agreement No. 551 715051; Spiders).

This work makes use of data supplied by the UK \emph{Swift} Science Data Centre at the University of Leicester and the \emph{Neil Gehrels Swift Observatory}.
This research has made use of data obtained from the Chandra Data Archive and software provided by the Chandra X-ray Center (CXC) in the application packages CIAO and Sherpa.

This work is based in part on observations taken by The Liverpool Telescope. The Liverpool Telescope is operated on the island of La Palma by Liverpool John Moores University in the Spanish Observatorio del Roque de los Muchachos of the Instituto de Astrofisica de Canarias with financial support from the UK Science and Technology Facilities Council.
The Gravitational-wave Optical Transient Observer (GOTO) project acknowledges the support of the Monash-Warwick Alliance; University of Warwick; Monash University; University of Sheffield; University of Leicester; Armagh Observatory \& Planetarium; the National Astronomical Research Institute of Thailand (NARIT); Instituto de Astrofísica de Canarias (IAC); University of Portsmouth; University of Turku, and the UK Science and Technology Facilities Council (STFC, grant numbers ST/T007184/1, ST/T003103/1).

\section*{Data Availability}
The data underlying this article will be shared on reasonable request to the corresponding author.

%%%%%%%%%%%%%%%%%%%%%%%%%%%%%%%%%%%%%%%%%%%%%%%%%%


\bibliographystyle{mnras}
\bibliography{ref}

\begin{thebibliography}{}
\makeatletter
\relax
\def\mn@urlcharsother{\let\do\@makeother \do\$\do\&\do\#\do\^\do\_\do\%\do\~}
\def\mn@doi{\begingroup\mn@urlcharsother \@ifnextchar [ {\mn@doi@}
  {\mn@doi@[]}}
\def\mn@doi@[#1]#2{\def\@tempa{#1}\ifx\@tempa\@empty \href
  {http://dx.doi.org/#2} {doi:#2}\else \href {http://dx.doi.org/#2} {#1}\fi
  \endgroup}
\def\mn@eprint#1#2{\mn@eprint@#1:#2::\@nil}
\def\mn@eprint@arXiv#1{\href {http://arxiv.org/abs/#1} {{\tt arXiv:#1}}}
\def\mn@eprint@dblp#1{\href {http://dblp.uni-trier.de/rec/bibtex/#1.xml}
  {dblp:#1}}
\def\mn@eprint@#1:#2:#3:#4\@nil{\def\@tempa {#1}\def\@tempb {#2}\def\@tempc
  {#3}\ifx \@tempc \@empty \let \@tempc \@tempb \let \@tempb \@tempa \fi \ifx
  \@tempb \@empty \def\@tempb {arXiv}\fi \@ifundefined
  {mn@eprint@\@tempb}{\@tempb:\@tempc}{\expandafter \expandafter \csname
  mn@eprint@\@tempb\endcsname \expandafter{\@tempc}}}

\bibitem[\protect\citeauthoryear{Abbott et~al.,}{Abbott
  et~al.}{2017}]{Abbott17a}
Abbott B.~P.,  et~al., 2017, \mn@doi [The Astrophysical Journal]
  {10.3847/2041-8213/aa920c}, 848, L13

\bibitem[\protect\citeauthoryear{{Ackley} et~al.,}{{Ackley}
  et~al.}{2020}]{Ackley20}
{Ackley} K.,  et~al., 2020, GRB Coordinates Network, \href
  {https://ui.adsabs.harvard.edu/abs/2020GCN.28639....1A} {28639, 1}

\bibitem[\protect\citeauthoryear{Ahumada et~al.,}{Ahumada
  et~al.}{2021}]{Ahumada21}
Ahumada T.,  et~al., 2021, \mn@doi [Nature Astronomy]
  {10.1038/s41550-021-01428-7}, 5, 917

\bibitem[\protect\citeauthoryear{{Amati}}{{Amati}}{2006}]{Amati06}
{Amati} L.,  2006, \mn@doi [\mnras] {10.1111/j.1365-2966.2006.10840.x}, \href
  {http://adsabs.harvard.edu/abs/2006MNRAS.372..233A} {372, 233}

\bibitem[\protect\citeauthoryear{Amati}{Amati}{2021}]{Amati21}
Amati L.,  2021, \mn@doi [Nature Astronomy] {10.1038/s41550-021-01401-4}, 5,
  877

\bibitem[\protect\citeauthoryear{{Amati} et~al.,}{{Amati}
  et~al.}{2002}]{Amati02}
{Amati} L.,  et~al., 2002, \mn@doi [\aap] {10.1051/0004-6361:20020722}, \href
  {http://adsabs.harvard.edu/abs/2002A%26A...390...81A} {390, 81}

\bibitem[\protect\citeauthoryear{{Amati} et~al.,}{{Amati}
  et~al.}{2021}]{Amati21b}
{Amati} L.,  et~al., 2021, \mn@doi [Experimental Astronomy]
  {10.1007/s10686-021-09807-8}, \href
  {https://ui.adsabs.harvard.edu/abs/2021ExA....52..183A} {52, 183}

\bibitem[\protect\citeauthoryear{{Arnaud}}{{Arnaud}}{1996}]{Arnaud96}
{Arnaud} K.~A.,  1996, in {Jacoby} G.~H.,  {Barnes} J.,  eds,  Astronomical
  Society of the Pacific Conference Series Vol. 101, Astronomical Data Analysis
  Software and Systems V. p.~17

\bibitem[\protect\citeauthoryear{{Band} et~al.,}{{Band} et~al.}{1993}]{Band93}
{Band} D.,  et~al., 1993, \mn@doi [\apj] {10.1086/172995}, \href
  {http://adsabs.harvard.edu/abs/1993ApJ...413..281B} {413, 281}

\bibitem[\protect\citeauthoryear{{Barbary}}{{Barbary}}{2016}]{Barbary16}
{Barbary} K.,  2016, \mn@doi [The Journal of Open Source Software]
  {10.21105/joss.00058}, \href
  {https://ui.adsabs.harvard.edu/abs/2016JOSS....1...58B} {1, 58}

\bibitem[\protect\citeauthoryear{{Barthelmy} et~al.,}{{Barthelmy}
  et~al.}{2005}]{Barthelmy05a}
{Barthelmy} S.~D.,  et~al., 2005, \mn@doi [\ssr] {10.1007/s11214-005-5096-3},
  \href {http://adsabs.harvard.edu/abs/2005SSRv..120..143B} {120, 143}

\bibitem[\protect\citeauthoryear{{Bazin} et~al.,}{{Bazin}
  et~al.}{2011}]{Bazin11}
{Bazin} G.,  et~al., 2011, \mn@doi [\aap] {10.1051/0004-6361/201116898}, \href
  {http://adsabs.harvard.edu/abs/2011A%26A...534A..43B} {534, A43}

\bibitem[\protect\citeauthoryear{{Beniamini} \& {van der Horst}}{{Beniamini} \&
  {van der Horst}}{2017}]{Beniamini17}
{Beniamini} P.,  {van der Horst} A.~J.,  2017, \mn@doi [\mnras]
  {10.1093/mnras/stx2203}, \href
  {http://adsabs.harvard.edu/abs/2017MNRAS.472.3161B} {472, 3161}

\bibitem[\protect\citeauthoryear{{Beniamini}, {Duque}, {Daigne}  \&
  {Mochkovitch}}{{Beniamini} et~al.}{2020}]{Beniamini20}
{Beniamini} P.,  {Duque} R.,  {Daigne} F.,   {Mochkovitch} R.,  2020, \mn@doi
  [\mnras] {10.1093/mnras/staa070}, \href
  {https://ui.adsabs.harvard.edu/abs/2020MNRAS.492.2847B} {492, 2847}

\bibitem[\protect\citeauthoryear{{Bernardini}, {Margutti}, {Mao}, {Zaninoni}
  \& {Chincarini}}{{Bernardini} et~al.}{2012}]{Bernardini12b}
{Bernardini} M.~G.,  {Margutti} R.,  {Mao} J.,  {Zaninoni} E.,   {Chincarini}
  G.,  2012, \mn@doi [\aap] {10.1051/0004-6361/201117895}, \href
  {https://ui.adsabs.harvard.edu/abs/2012A&A...539A...3B} {539, A3}

\bibitem[\protect\citeauthoryear{Bernardini, Cordier  \& Wei}{Bernardini
  et~al.}{2021}]{Bernardini21}
Bernardini M.~G.,  Cordier B.,   Wei J.,  2021, \mn@doi [Galaxies]
  {10.3390/galaxies9040113}, 9

\bibitem[\protect\citeauthoryear{Berti \& Carosi}{Berti \&
  Carosi}{2022}]{Berti22}
Berti A.,  Carosi A.,  2022, \mn@doi [Galaxies] {10.3390/galaxies10030067}, 10

\bibitem[\protect\citeauthoryear{{Bertin} \& {Arnouts}}{{Bertin} \&
  {Arnouts}}{1996}]{Bertin96}
{Bertin} E.,  {Arnouts} S.,  1996, \mn@doi [\aaps] {10.1051/aas:1996164}, \href
  {https://ui.adsabs.harvard.edu/abs/1996A&AS..117..393B} {117, 393}

\bibitem[\protect\citeauthoryear{{Blanch} et~al.,}{{Blanch}
  et~al.}{2020}]{GCN28659}
{Blanch} O.,  et~al., 2020, GRB Coordinates Network, \href
  {https://ui.adsabs.harvard.edu/abs/2020GCN.28659....1B} {28659, 1}

\bibitem[\protect\citeauthoryear{{Burrows} et~al.,}{{Burrows}
  et~al.}{2005}]{Burrows05}
{Burrows} D.~N.,  et~al., 2005, \mn@doi [Science] {10.1126/science.1116168},
  \href {http://adsabs.harvard.edu/abs/2005Sci...309.1833B} {309, 1833}

\bibitem[\protect\citeauthoryear{Campana et~al.,}{Campana
  et~al.}{2006}]{Campana06}
Campana S.,  et~al., 2006, \mn@doi [Nature] {10.1038/nature04892}, 442, 1008

\bibitem[\protect\citeauthoryear{{Cash}}{{Cash}}{1979}]{Cash79}
{Cash} W.,  1979, \mn@doi [\apj] {10.1086/156922}, \href
  {https://ui.adsabs.harvard.edu/abs/1979ApJ...228..939C} {228, 939}

\bibitem[\protect\citeauthoryear{{Chambers} et~al.,}{{Chambers}
  et~al.}{2016}]{Chambers16}
{Chambers} K.~C.,  et~al., 2016, arXiv e-prints, \href
  {https://ui.adsabs.harvard.edu/abs/2016arXiv161205560C} {p. arXiv:1612.05560}

\bibitem[\protect\citeauthoryear{{Chen}, {Xie}, {Lei}, {Zou}, {L{\"u}},
  {Liang}, {Gao}  \& {Wang}}{{Chen} et~al.}{2017}]{Chen17}
{Chen} W.,  {Xie} W.,  {Lei} W.-H.,  {Zou} Y.-C.,  {L{\"u}} H.-J.,  {Liang}
  E.-W.,  {Gao} H.,   {Wang} D.-X.,  2017, \mn@doi [\apj]
  {10.3847/1538-4357/aa8f4a}, \href
  {https://ui.adsabs.harvard.edu/abs/2017ApJ...849..119C} {849, 119}

\bibitem[\protect\citeauthoryear{Chen, Urata  \& Huang}{Chen
  et~al.}{2021}]{Chen21}
Chen J.-C.,  Urata Y.,   Huang K.,  2021, \mn@doi [The Astrophysical Journal]
  {10.3847/1538-4357/ac00b4}, 915, 46

\bibitem[\protect\citeauthoryear{{Chevalier} \& {Li}}{{Chevalier} \&
  {Li}}{1999}]{Chevalier99}
{Chevalier} R.~A.,  {Li} Z.-Y.,  1999, \mn@doi [\apjl] {10.1086/312147}, \href
  {http://adsabs.harvard.edu/abs/1999ApJ...520L..29C} {520, L29}

\bibitem[\protect\citeauthoryear{D'Avanzo et~al.,}{D'Avanzo
  et~al.}{2014}]{D'Avanzo14}
D'Avanzo P.,  et~al., 2014, \mn@doi [Monthly Notices of the Royal Astronomical
  Society] {10.1093/mnras/stu994}, 442, 2342

\bibitem[\protect\citeauthoryear{{D'Elia} et~al.,}{{D'Elia}
  et~al.}{2020}]{GCN28632}
{D'Elia} V.,  et~al., 2020, GRB Coordinates Network, \href
  {https://ui.adsabs.harvard.edu/abs/2020GCN.28632....1D} {28632, 1}

\bibitem[\protect\citeauthoryear{Dainotti, Petrosian, Willingale, O'Brien,
  Ostrowski  \& Nagataki}{Dainotti et~al.}{2015}]{Dainotti15}
Dainotti M.,  Petrosian V.,  Willingale R.,  O'Brien P.,  Ostrowski M.,
  Nagataki S.,  2015, \mn@doi [Monthly Notices of the Royal Astronomical
  Society] {10.1093/mnras/stv1229}, 451, 3898

\bibitem[\protect\citeauthoryear{{Dainotti}, {Hernandez}, {Postnikov},
  {Nagataki}, {O'brien}, {Willingale}  \& {Striegel}}{{Dainotti}
  et~al.}{2017}]{Dainotti17}
{Dainotti} M.~G.,  {Hernandez} X.,  {Postnikov} S.,  {Nagataki} S.,  {O'brien}
  P.,  {Willingale} R.,   {Striegel} S.,  2017, \mn@doi [\apj]
  {10.3847/1538-4357/aa8a6b}, \href
  {https://ui.adsabs.harvard.edu/abs/2017ApJ...848...88D} {848, 88}

\bibitem[\protect\citeauthoryear{{Evans} et~al.,}{{Evans}
  et~al.}{2007}]{Evans07}
{Evans} P.~A.,  et~al., 2007, \mn@doi [\aap] {10.1051/0004-6361:20077530},
  \href {http://adsabs.harvard.edu/abs/2007A%26A...469..379E} {469, 379}

\bibitem[\protect\citeauthoryear{{Evans} et~al.,}{{Evans}
  et~al.}{2009}]{Evans09}
{Evans} P.~A.,  et~al., 2009, \mn@doi [\mnras]
  {10.1111/j.1365-2966.2009.14913.x}, \href
  {http://adsabs.harvard.edu/abs/2009MNRAS.397.1177E} {397, 1177}

\bibitem[\protect\citeauthoryear{{Evans} et~al.,}{{Evans}
  et~al.}{2010}]{Evans10}
{Evans} P.~A.,  et~al., 2010, \mn@doi [\aap] {10.1051/0004-6361/201014819},
  \href {https://ui.adsabs.harvard.edu/abs/2010A&A...519A.102E} {519, A102}

\bibitem[\protect\citeauthoryear{{Evans}, {Goad}, {Osborne}, {Beardmore}  \&
  {Swift-XRT Team.}}{{Evans} et~al.}{2020}]{GCN28647}
{Evans} P.~A.,  {Goad} M.~R.,  {Osborne} J.~P.,  {Beardmore} A.~P.,
  {Swift-XRT Team.} 2020, GRB Coordinates Network, \href
  {https://ui.adsabs.harvard.edu/abs/2020GCN.28647....1E} {28647, 1}

\bibitem[\protect\citeauthoryear{Fargion}{Fargion}{2006}]{GCN4819}
Fargion D.,  2006, GRB Coordinates Network, 4819

\bibitem[\protect\citeauthoryear{{Fletcher}, {Veres}  \& {Fermi-GBM
  Team}}{{Fletcher} et~al.}{2020}]{GCN28663}
{Fletcher} C.,  {Veres} P.,   {Fermi-GBM Team} 2020, GRB Coordinates Network,
  \href {https://ui.adsabs.harvard.edu/abs/2020GCN.28663....1F} {28663, 1}

\bibitem[\protect\citeauthoryear{{Fong} et~al.,}{{Fong} et~al.}{2014}]{Fong14}
{Fong} W.,  et~al., 2014, \mn@doi [\apj] {10.1088/0004-637X/780/2/118}, \href
  {http://adsabs.harvard.edu/abs/2014ApJ...780..118F} {780, 118}

\bibitem[\protect\citeauthoryear{{Fong}, {Schroeder}, {Rastinejad}  \&
  {Hajela}}{{Fong} et~al.}{2020}]{GCN28688}
{Fong} W.,  {Schroeder} G.,  {Rastinejad} J.,   {Hajela} A.,  2020, GRB
  Coordinates Network, \href
  {https://ui.adsabs.harvard.edu/abs/2020GCN.28688....1F} {28688, 1}

\bibitem[\protect\citeauthoryear{{Fong} et~al.,}{{Fong} et~al.}{2022}]{Fong22}
{Fong} W.-f.,  et~al., 2022, arXiv e-prints, \href
  {https://ui.adsabs.harvard.edu/abs/2022arXiv220601763F} {p. arXiv:2206.01763}

\bibitem[\protect\citeauthoryear{{Frei} \& {Gunn}}{{Frei} \&
  {Gunn}}{1994}]{Frei1994}
{Frei} Z.,  {Gunn} J.~E.,  1994, \mn@doi [\aj] {10.1086/117172}, \href
  {https://ui.adsabs.harvard.edu/abs/1994AJ....108.1476F} {108, 1476}

\bibitem[\protect\citeauthoryear{{Fruchter} et~al.,}{{Fruchter}
  et~al.}{2006}]{Fruchter06}
{Fruchter} A.~S.,  et~al., 2006, \mn@doi [\nat] {10.1038/nature04787}, \href
  {http://adsabs.harvard.edu/abs/2006Natur.441..463F} {441, 463}

\bibitem[\protect\citeauthoryear{{Fukugita}, {Ichikawa}, {Gunn}, {Doi},
  {Shimasaku}  \& {Schneider}}{{Fukugita} et~al.}{1996}]{Fukugita96}
{Fukugita} M.,  {Ichikawa} T.,  {Gunn} J.~E.,  {Doi} M.,  {Shimasaku} K.,
  {Schneider} D.~P.,  1996, \mn@doi [\aj] {10.1086/117915}, \href
  {https://ui.adsabs.harvard.edu/abs/1996AJ....111.1748F} {111, 1748}

\bibitem[\protect\citeauthoryear{{Galama} et~al.,}{{Galama}
  et~al.}{1999}]{Galama99}
{Galama} T.~J.,  et~al., 1999, \mn@doi [\aaps] {10.1051/aas:1999311}, \href
  {https://ui.adsabs.harvard.edu/abs/1999A&AS..138..465G} {138, 465}

\bibitem[\protect\citeauthoryear{{Gao}, {Lei}, {Zou}, {Wu}  \& {Zhang}}{{Gao}
  et~al.}{2013a}]{Gao13}
{Gao} H.,  {Lei} W.-H.,  {Zou} Y.-C.,  {Wu} X.-F.,   {Zhang} B.,  2013a,
  \mn@doi [\nar] {10.1016/j.newar.2013.10.001}, \href
  {http://adsabs.harvard.edu/abs/2013NewAR..57..141G} {57, 141}

\bibitem[\protect\citeauthoryear{{Gao}, {Ding}, {Wu}, {Zhang}  \& {Dai}}{{Gao}
  et~al.}{2013b}]{Gao13b}
{Gao} H.,  {Ding} X.,  {Wu} X.-F.,  {Zhang} B.,   {Dai} Z.-G.,  2013b, \mn@doi
  [\apj] {10.1088/0004-637X/771/2/86}, \href
  {http://adsabs.harvard.edu/abs/2013ApJ...771...86G} {771, 86}

\bibitem[\protect\citeauthoryear{{Gehrels} et~al.,}{{Gehrels}
  et~al.}{2004}]{Gehrels04}
{Gehrels} N.,  et~al., 2004, \mn@doi [\apj] {10.1086/422091}, \href
  {http://adsabs.harvard.edu/abs/2004ApJ...611.1005G} {611, 1005}

\bibitem[\protect\citeauthoryear{{Gehrels} et~al.,}{{Gehrels}
  et~al.}{2006}]{Gehrels06}
{Gehrels} N.,  et~al., 2006, \mn@doi [\nat] {10.1038/nature05376}, \href
  {http://adsabs.harvard.edu/abs/2006Natur.444.1044G} {444, 1044}

\bibitem[\protect\citeauthoryear{{Gehrels} et~al.,}{{Gehrels}
  et~al.}{2008}]{Gehrels08}
{Gehrels} N.,  et~al., 2008, \mn@doi [\apj] {10.1086/592766}, \href
  {http://adsabs.harvard.edu/abs/2008ApJ...689.1161G} {689, 1161}

\bibitem[\protect\citeauthoryear{Giarratana et~al.,}{Giarratana
  et~al.}{2022}]{Giarratana22}
Giarratana S.,  et~al., 2022, arXiv preprint arXiv:2205.12750

\bibitem[\protect\citeauthoryear{{Goad} et~al.,}{{Goad} et~al.}{2007}]{Goad07}
{Goad} M.~R.,  et~al., 2007, \mn@doi [\aap] {10.1051/0004-6361:20078436}, \href
  {http://adsabs.harvard.edu/abs/2007A%26A...476.1401G} {476, 1401}

\bibitem[\protect\citeauthoryear{{Gompertz}, {Fruchter}  \& {Pe'er}}{{Gompertz}
  et~al.}{2018}]{Gompertz18b}
{Gompertz} B.~P.,  {Fruchter} A.~S.,   {Pe'er} A.,  2018, \mn@doi [\apj]
  {10.3847/1538-4357/aadba8}, \href
  {http://adsabs.harvard.edu/abs/2018ApJ...866..162G} {866, 162}

\bibitem[\protect\citeauthoryear{{Gompertz}, {Levan}  \& {Tanvir}}{{Gompertz}
  et~al.}{2020}]{Gompertz20}
{Gompertz} B.~P.,  {Levan} A.~J.,   {Tanvir} N.~R.,  2020, \mn@doi [\apj]
  {10.3847/1538-4357/ab8d24}, \href
  {https://ui.adsabs.harvard.edu/abs/2020ApJ...895...58G} {895, 58}

\bibitem[\protect\citeauthoryear{{Gompertz} et~al.,}{{Gompertz}
  et~al.}{2022}]{Gompertz22}
{Gompertz} B.~P.,  et~al., 2022, arXiv e-prints, \href
  {https://ui.adsabs.harvard.edu/abs/2022arXiv220505008G} {p. arXiv:2205.05008}

\bibitem[\protect\citeauthoryear{{Greiner} et~al.,}{{Greiner}
  et~al.}{2015}]{Greiner15}
{Greiner} J.,  et~al., 2015, \mn@doi [\nat] {10.1038/nature14579}, \href
  {http://adsabs.harvard.edu/abs/2015Natur.523..189G} {523, 189}

\bibitem[\protect\citeauthoryear{{Hjorth} et~al.,}{{Hjorth}
  et~al.}{2003}]{Hjorth03}
{Hjorth} J.,  et~al., 2003, \mn@doi [\nat] {10.1038/nature01750}, \href
  {http://adsabs.harvard.edu/abs/2003Natur.423..847H} {423, 847}

\bibitem[\protect\citeauthoryear{Irwin \& Chevalier}{Irwin \&
  Chevalier}{2016}]{Irwin16}
Irwin C.~M.,  Chevalier R.~A.,  2016, \mn@doi [Monthly Notices of the Royal
  Astronomical Society] {10.1093/mnras/stw1058}, 460, 1680

\bibitem[\protect\citeauthoryear{{Iyyani}, {Ryde}, {Burgess}, {Pe'er}  \&
  {B{\'e}gu{\'e}}}{{Iyyani} et~al.}{2016}]{Iyyani16}
{Iyyani} S.,  {Ryde} F.,  {Burgess} J.~M.,  {Pe'er} A.,   {B{\'e}gu{\'e}} D.,
  2016, \mn@doi [\mnras] {10.1093/mnras/stv2751}, \href
  {https://ui.adsabs.harvard.edu/abs/2016MNRAS.456.2157I} {456, 2157}

\bibitem[\protect\citeauthoryear{Jespersen, Severin, Steinhardt, Vinther,
  Fynbo, Selsing  \& Watson}{Jespersen et~al.}{2020}]{Jespersen20}
Jespersen C.~K.,  Severin J.~B.,  Steinhardt C.~L.,  Vinther J.,  Fynbo J.
  P.~U.,  Selsing J.,   Watson D.,  2020, \mn@doi [The Astrophysical Journal]
  {10.3847/2041-8213/ab964d}, 896, L20

\bibitem[\protect\citeauthoryear{{Kann} et~al.,}{{Kann} et~al.}{2016}]{Kann16}
{Kann} D.~A.,  et~al., 2016, preprint, \href
  {http://adsabs.harvard.edu/abs/2016arXiv160606791K} {} (\mn@eprint {arXiv}
  {1606.06791})

\bibitem[\protect\citeauthoryear{{Klebesadel}, {Strong}  \&
  {Olson}}{{Klebesadel} et~al.}{1973}]{Klebesadel73}
{Klebesadel} R.~W.,  {Strong} I.~B.,   {Olson} R.~A.,  1973, \mn@doi [\apjl]
  {10.1086/181225}, \href {http://adsabs.harvard.edu/abs/1973ApJ...182L..85K}
  {182, L85}

\bibitem[\protect\citeauthoryear{Klu{\'{z}}niak \& Ruderman}{Klu{\'{z}}niak \&
  Ruderman}{1998}]{Kluzniak98}
Klu{\'{z}}niak W.,  Ruderman M.,  1998, \mn@doi [The Astrophysical Journal]
  {10.1086/311622}, 505, L113

\bibitem[\protect\citeauthoryear{{Koshut}, {Paciesas}, {Kouveliotou}, {van
  Paradijs}, {Pendleton}, {Fishman}  \& {Meegan}}{{Koshut}
  et~al.}{1995}]{Koshut95}
{Koshut} T.~M.,  {Paciesas} W.~S.,  {Kouveliotou} C.,  {van Paradijs} J.,
  {Pendleton} G.~N.,  {Fishman} G.~J.,   {Meegan} C.~A.,  1995, in American
  Astronomical Society Meeting Abstracts \#186. p. 53.01

\bibitem[\protect\citeauthoryear{{Kouveliotou}, {Meegan}, {Fishman}, {Bhat},
  {Briggs}, {Koshut}, {Paciesas}  \& {Pendleton}}{{Kouveliotou}
  et~al.}{1993}]{Kouveliotou93}
{Kouveliotou} C.,  {Meegan} C.~A.,  {Fishman} G.~J.,  {Bhat} N.~P.,  {Briggs}
  M.~S.,  {Koshut} T.~M.,  {Paciesas} W.~S.,   {Pendleton} G.~N.,  1993,
  \mn@doi [\apjl] {10.1086/186969}, \href
  {http://adsabs.harvard.edu/abs/1993ApJ...413L.101K} {413, L101}

\bibitem[\protect\citeauthoryear{{Lamb} et~al.,}{{Lamb} et~al.}{2019}]{Lamb19b}
{Lamb} G.~P.,  et~al., 2019, \mn@doi [\apj] {10.3847/1538-4357/ab38bb}, \href
  {https://ui.adsabs.harvard.edu/abs/2019ApJ...883...48L} {883, 48}

\bibitem[\protect\citeauthoryear{Lamb, Levan  \& Tanvir}{Lamb
  et~al.}{2020}]{Lamb20}
Lamb G.~P.,  Levan A.~J.,   Tanvir N.~R.,  2020, \mn@doi [The Astrophysical
  Journal] {10.3847/1538-4357/aba75a}, 899, 105

\bibitem[\protect\citeauthoryear{{Lamb} et~al.,}{{Lamb}
  et~al.}{2021a}]{Lamb21b}
{Lamb} G.~P.,  et~al., 2021a, \mn@doi [Universe] {10.3390/universe7090329},
  \href {https://ui.adsabs.harvard.edu/abs/2021Univ....7..329L} {7, 329}

\bibitem[\protect\citeauthoryear{Lamb, Kann, Fernández, Mandel, Levan  \&
  Tanvir}{Lamb et~al.}{2021b}]{Lamb21}
Lamb G.~P.,  Kann D.~A.,  Fernández J.~J.,  Mandel I.,  Levan A.~J.,   Tanvir
  N.~R.,  2021b, \mn@doi [Monthly Notices of the Royal Astronomical Society]
  {10.1093/mnras/stab2071}, 506, 4163

\bibitem[\protect\citeauthoryear{{Lang}, {Hogg}, {Mierle}, {Blanton}  \&
  {Roweis}}{{Lang} et~al.}{2010}]{Lang10}
{Lang} D.,  {Hogg} D.~W.,  {Mierle} K.,  {Blanton} M.,   {Roweis} S.,  2010,
  \mn@doi [\aj] {10.1088/0004-6256/139/5/1782}, \href
  {https://ui.adsabs.harvard.edu/abs/2010AJ....139.1782L} {139, 1782}

\bibitem[\protect\citeauthoryear{Le \& Mehta}{Le \& Mehta}{2017}]{Le17}
Le T.,  Mehta V.,  2017, \mn@doi [The Astrophysical Journal]
  {10.3847/1538-4357/aa5fa7}, 837, 17

\bibitem[\protect\citeauthoryear{{Lei}, {Zhang}  \& {Liang}}{{Lei}
  et~al.}{2013}]{Lei13}
{Lei} W.-H.,  {Zhang} B.,   {Liang} E.-W.,  2013, \mn@doi [\apj]
  {10.1088/0004-637X/765/2/125}, \href
  {http://adsabs.harvard.edu/abs/2013ApJ...765..125L} {765, 125}

\bibitem[\protect\citeauthoryear{Levan}{Levan}{2018}]{Levan18}
Levan A.,  2018, in 2514-3433, Gamma-Ray Bursts.
IOP Publishing, pp 5--1 to 5--18, \mn@doi{10.1088/2514-3433/aae164ch5}, \url
  {https://dx.doi.org/10.1088/2514-3433/aae164ch5}

\bibitem[\protect\citeauthoryear{{Levan} et~al.,}{{Levan}
  et~al.}{2014}]{Levan14}
{Levan} A.~J.,  et~al., 2014, \mn@doi [\apj] {10.1088/0004-637X/781/1/13},
  \href {http://adsabs.harvard.edu/abs/2014ApJ...781...13L} {781, 13}

\bibitem[\protect\citeauthoryear{{Levan}, {Crowther}, {de Grijs}, {Langer},
  {Xu}  \& {Yoon}}{{Levan} et~al.}{2016}]{Levan16}
{Levan} A.,  {Crowther} P.,  {de Grijs} R.,  {Langer} N.,  {Xu} D.,   {Yoon}
  S.-C.,  2016, \mn@doi [\ssr] {10.1007/s11214-016-0312-x}, \href
  {http://adsabs.harvard.edu/abs/2016SSRv..202...33L} {202, 33}

\bibitem[\protect\citeauthoryear{Li, Wu, Lei, Dai, Liang  \& Ryde}{Li
  et~al.}{2018}]{Li18}
Li L.,  Wu X.-F.,  Lei W.-H.,  Dai Z.-G.,  Liang E.-W.,   Ryde F.,  2018,
  \mn@doi [The Astrophysical Journal Supplement Series]
  {10.3847/1538-4365/aabaf3}, 236, 26

\bibitem[\protect\citeauthoryear{Li et~al.,}{Li et~al.}{2020}]{Li20}
Li L.,  et~al., 2020, \mn@doi [The Astrophysical Journal]
  {10.3847/1538-4357/aba757}, 900, 176

\bibitem[\protect\citeauthoryear{Liang, Zhang, Virgili  \& Dai}{Liang
  et~al.}{2007}]{Liang07}
Liang E.,  Zhang B.,  Virgili F.,   Dai Z.~G.,  2007, \mn@doi [The
  Astrophysical Journal] {10.1086/517959}, 662, 1111

\bibitem[\protect\citeauthoryear{{Lien} et~al.,}{{Lien} et~al.}{2016}]{Lien16}
{Lien} A.,  et~al., 2016, \mn@doi [\apj] {10.3847/0004-637X/829/1/7}, \href
  {http://adsabs.harvard.edu/abs/2016ApJ...829....7L} {829, 7}

\bibitem[\protect\citeauthoryear{{Lipunov} et~al.,}{{Lipunov}
  et~al.}{2010}]{Lipunov10}
{Lipunov} V.,  et~al., 2010, \mn@doi [Advances in Astronomy]
  {10.1155/2010/349171}, \href
  {https://ui.adsabs.harvard.edu/abs/2010AdAst2010E..30L} {2010, 349171}

\bibitem[\protect\citeauthoryear{{Lipunov} et~al.,}{{Lipunov}
  et~al.}{2020}]{Lipunov20}
{Lipunov} V.,  et~al., 2020, GRB Coordinates Network, \href
  {https://ui.adsabs.harvard.edu/abs/2020GCN.28633....1L} {28633, 1}

\bibitem[\protect\citeauthoryear{{L{\"u}}, {Zhang}, {Lei}, {Li}  \&
  {Lasky}}{{L{\"u}} et~al.}{2015}]{Lu15}
{L{\"u}} H.-J.,  {Zhang} B.,  {Lei} W.-H.,  {Li} Y.,   {Lasky} P.~D.,  2015,
  preprint, \href {http://adsabs.harvard.edu/abs/2015arXiv150102589L}
  {1501.02589}

\bibitem[\protect\citeauthoryear{Lyman, Bersier  \& James}{Lyman
  et~al.}{2014}]{Lyman14}
Lyman J.~D.,  Bersier D.,   James P.~A.,  2014, \mn@doi [Monthly Notices of the
  Royal Astronomical Society] {10.1093/mnras/stt2187}, 437, 3848

\bibitem[\protect\citeauthoryear{Lyman, Bersier, James, Mazzali, Eldridge,
  Fraser  \& Pian}{Lyman et~al.}{2016}]{Lyman16}
Lyman J.~D.,  Bersier D.,  James P.~A.,  Mazzali P.~A.,  Eldridge J.~J.,
  Fraser M.,   Pian E.,  2016, \mn@doi [Monthly Notices of the Royal
  Astronomical Society] {10.1093/mnras/stv2983}, 457, 328

\bibitem[\protect\citeauthoryear{{Malesani}, {de Ugarte Postigo}  \&
  {Pursimo}}{{Malesani} et~al.}{2020}]{Malesani20}
{Malesani} D.~B.,  {de Ugarte Postigo} A.,   {Pursimo} T.,  2020, GRB
  Coordinates Network, \href
  {https://ui.adsabs.harvard.edu/abs/2020GCN.28637....1M} {28637, 1}

\bibitem[\protect\citeauthoryear{{Margutti} et~al.,}{{Margutti}
  et~al.}{2013}]{Margutti13}
{Margutti} R.,  et~al., 2013, \mn@doi [\mnras] {10.1093/mnras/sts066}, \href
  {https://ui.adsabs.harvard.edu/abs/2013MNRAS.428..729M} {428, 729}

\bibitem[\protect\citeauthoryear{{Markwardt} et~al.,}{{Markwardt}
  et~al.}{2020}]{GCN28658}
{Markwardt} C.~B.,  et~al., 2020, GRB Coordinates Network, \href
  {https://ui.adsabs.harvard.edu/abs/2020GCN.28658....1M} {28658, 1}

\bibitem[\protect\citeauthoryear{{Meegan} et~al.,}{{Meegan}
  et~al.}{2009}]{Meegan09}
{Meegan} C.,  et~al., 2009, \mn@doi [\apj] {10.1088/0004-637X/702/1/791}, \href
  {http://adsabs.harvard.edu/abs/2009ApJ...702..791M} {702, 791}

\bibitem[\protect\citeauthoryear{{M{\'e}sz{\'a}ros}}{{M{\'e}sz{\'a}ros}}{2002}]{Meszaros02}
{M{\'e}sz{\'a}ros} P.,  2002, \mn@doi [\araa]
  {10.1146/annurev.astro.40.060401.093821}, \href
  {http://adsabs.harvard.edu/abs/2002ARA%26A..40..137M} {40, 137}

\bibitem[\protect\citeauthoryear{{Minaev} \& {Pozanenko}}{{Minaev} \&
  {Pozanenko}}{2020a}]{Minaev20}
{Minaev} P.~Y.,  {Pozanenko} A.~S.,  2020a, \mn@doi [\mnras]
  {10.1093/mnras/stz3611}, \href
  {https://ui.adsabs.harvard.edu/abs/2020MNRAS.492.1919M} {492, 1919}

\bibitem[\protect\citeauthoryear{{Minaev} \& {Pozanenko}}{{Minaev} \&
  {Pozanenko}}{2020b}]{GCN28668}
{Minaev} P.,  {Pozanenko} A.,  2020b, GRB Coordinates Network, \href
  {https://ui.adsabs.harvard.edu/abs/2020GCN.28668....1M} {28668, 1}

\bibitem[\protect\citeauthoryear{{Mukherjee}, {Feigelson}, {Jogesh Babu},
  {Murtagh}, {Fraley}  \& {Raftery}}{{Mukherjee} et~al.}{1998}]{Mukherjee98}
{Mukherjee} S.,  {Feigelson} E.~D.,  {Jogesh Babu} G.,  {Murtagh} F.,  {Fraley}
  C.,   {Raftery} A.,  1998, \mn@doi [\apj] {10.1086/306386}, \href
  {https://ui.adsabs.harvard.edu/abs/1998ApJ...508..314M} {508, 314}

\bibitem[\protect\citeauthoryear{N.~Masetti, Pian  \& Patat}{N.~Masetti
  et~al.}{2006}]{GCN4803}
N.~Masetti E.~P.,  Pian E.,   Patat F.,  2006, GRB Coordinates Network, 4803

\bibitem[\protect\citeauthoryear{{Nakar}}{{Nakar}}{2015}]{Nakar15}
{Nakar} E.,  2015, \mn@doi [\apj] {10.1088/0004-637X/807/2/172}, \href
  {https://ui.adsabs.harvard.edu/abs/2015ApJ...807..172N} {807, 172}

\bibitem[\protect\citeauthoryear{{Nava}}{{Nava}}{2018}]{Nava18}
{Nava} L.,  2018, \mn@doi [International Journal of Modern Physics D]
  {10.1142/S0218271818420038}, \href
  {https://ui.adsabs.harvard.edu/abs/2018IJMPD..2742003N} {27, 1842003}

\bibitem[\protect\citeauthoryear{{Norris} \& {Bonnell}}{{Norris} \&
  {Bonnell}}{2006}]{Norris06}
{Norris} J.~P.,  {Bonnell} J.~T.,  2006, \mn@doi [\apj] {10.1086/502796}, \href
  {http://adsabs.harvard.edu/abs/2006ApJ...643..266N} {643, 266}

\bibitem[\protect\citeauthoryear{{Norris}, {Gehrels}  \& {Scargle}}{{Norris}
  et~al.}{2010}]{Norris10}
{Norris} J.~P.,  {Gehrels} N.,   {Scargle} J.~D.,  2010, \mn@doi [\apj]
  {10.1088/0004-637X/717/1/411}, \href
  {http://adsabs.harvard.edu/abs/2010ApJ...717..411N} {717, 411}

\bibitem[\protect\citeauthoryear{{Nousek} et~al.,}{{Nousek}
  et~al.}{2006}]{Nousek06}
{Nousek} J.~A.,  et~al., 2006, \mn@doi [\apj] {10.1086/500724}, \href
  {http://adsabs.harvard.edu/abs/2006ApJ...642..389N} {642, 389}

\bibitem[\protect\citeauthoryear{{Panaitescu} \& {Kumar}}{{Panaitescu} \&
  {Kumar}}{2002}]{Panaitescu02}
{Panaitescu} A.,  {Kumar} P.,  2002, \mn@doi [\apj] {10.1086/340094}, \href
  {http://adsabs.harvard.edu/abs/2002ApJ...571..779P} {571, 779}

\bibitem[\protect\citeauthoryear{Peng, Königl  \& Granot}{Peng
  et~al.}{2005}]{Peng05}
Peng F.,  Königl A.,   Granot J.,  2005, \mn@doi [The Astrophysical Journal]
  {10.1086/430045}, 626, 966

\bibitem[\protect\citeauthoryear{Piran}{Piran}{2005}]{Piran04}
Piran T.,  2005, \mn@doi [Rev. Mod. Phys.] {10.1103/RevModPhys.76.1143}, 76,
  1143

\bibitem[\protect\citeauthoryear{Prentice et~al.,}{Prentice
  et~al.}{2019}]{Prentice19}
Prentice S.~J.,  et~al., 2019, \mn@doi [Monthly Notices of the Royal
  Astronomical Society] {10.1093/mnras/sty3399}, 485, 1559

\bibitem[\protect\citeauthoryear{{Racusin} et~al.,}{{Racusin}
  et~al.}{2009}]{Racusin09b}
{Racusin} J.~L.,  et~al., 2009, \mn@doi [\apj] {10.1088/0004-637X/698/1/43},
  \href {http://adsabs.harvard.edu/abs/2009ApJ...698...43R} {698, 43}

\bibitem[\protect\citeauthoryear{{Rastinejad}, {Paterson}, {Kilpatrick}  \&
  {Fong}}{{Rastinejad} et~al.}{2020}]{GCN28676}
{Rastinejad} J.,  {Paterson} K.,  {Kilpatrick} C.~D.,   {Fong} W.,  2020, GRB
  Coordinates Network, \href
  {https://ui.adsabs.harvard.edu/abs/2020GCN.28676....1R} {28676, 1}

\bibitem[\protect\citeauthoryear{{Rastinejad} et~al.,}{{Rastinejad}
  et~al.}{2022}]{Rastinejad22}
{Rastinejad} J.~C.,  et~al., 2022, arXiv e-prints, \href
  {https://ui.adsabs.harvard.edu/abs/2022arXiv220410864R} {p. arXiv:2204.10864}

\bibitem[\protect\citeauthoryear{{Rhodes}, {Fender}, {Bray}  \&
  {Williams}}{{Rhodes} et~al.}{2020}]{GCN28945}
{Rhodes} L.,  {Fender} R.,  {Bray} J.,   {Williams} D.~R.~A.,  2020, GRB
  Coordinates Network, \href
  {https://ui.adsabs.harvard.edu/abs/2020GCN.28945....1R} {28945, 1}

\bibitem[\protect\citeauthoryear{Ricker, Atteia, Kawai, Lamb  \&
  Woosley}{Ricker et~al.}{2002}]{GCN1530}
Ricker G.,  Atteia J.-L.,  Kawai N.,  Lamb D.,   Woosley S.,  2002, GRB
  Coordinates Network, 1530

\bibitem[\protect\citeauthoryear{{Rossi}, {Benetti}, {Palazzi}, {D'Avanzo},
  {D'Elia}, {De Pasquale}  \& {CIBO Collaboration}}{{Rossi}
  et~al.}{2021}]{GCN29306}
{Rossi} A.,  {Benetti} S.,  {Palazzi} E.,  {D'Avanzo} P.,  {D'Elia} V.,  {De
  Pasquale} M.,   {CIBO Collaboration} 2021, GRB Coordinates Network, \href
  {https://ui.adsabs.harvard.edu/abs/2021GCN.29306....1R} {29306, 1}

\bibitem[\protect\citeauthoryear{Sakamoto et~al.,}{Sakamoto
  et~al.}{2014}]{GCN16438}
Sakamoto T.,  et~al., 2014, GRB Coordinates Network, 16438

\bibitem[\protect\citeauthoryear{{Sari}, {Piran}  \& {Narayan}}{{Sari}
  et~al.}{1998}]{Sari98}
{Sari} R.,  {Piran} T.,   {Narayan} R.,  1998, \mn@doi [\apjl]
  {10.1086/311269}, \href {http://adsabs.harvard.edu/abs/1998ApJ...497L..17S}
  {497, L17}

\bibitem[\protect\citeauthoryear{{Sari}, {Piran}  \& {Halpern}}{{Sari}
  et~al.}{1999}]{Sari99}
{Sari} R.,  {Piran} T.,   {Halpern} J.~P.,  1999, \mn@doi [\apjl]
  {10.1086/312109}, \href {http://adsabs.harvard.edu/abs/1999ApJ...519L..17S}
  {519, L17}

\bibitem[\protect\citeauthoryear{Sato, Obayashi, Yamazaki, Murase  \&
  Ohira}{Sato et~al.}{2021}]{Sato21}
Sato Y.,  Obayashi K.,  Yamazaki R.,  Murase K.,   Ohira Y.,  2021, \mn@doi
  [Monthly Notices of the Royal Astronomical Society] {10.1093/mnras/stab1273},
  504, 5647

\bibitem[\protect\citeauthoryear{{Schlafly} \& {Finkbeiner}}{{Schlafly} \&
  {Finkbeiner}}{2011}]{Schlafly11}
{Schlafly} E.~F.,  {Finkbeiner} D.~P.,  2011, \mn@doi [\apj]
  {10.1088/0004-637X/737/2/103}, \href
  {http://adsabs.harvard.edu/abs/2011ApJ...737..103S} {737, 103}

\bibitem[\protect\citeauthoryear{{Schulze} et~al.,}{{Schulze}
  et~al.}{2014}]{Schulze14a}
{Schulze} S.,  et~al., 2014, \mn@doi [\aap] {10.1051/0004-6361/201423387},
  \href {https://ui.adsabs.harvard.edu/abs/2014A&A...566A.102S} {566, A102}

\bibitem[\protect\citeauthoryear{Soderberg, Price, Fox, Kulkarni, Djorgovski,
  Berger, Harrison  \& Yost}{Soderberg et~al.}{2002}]{GCN1554}
Soderberg A.,  Price P.,  Fox D.,  Kulkarni S.,  Djorgovski S.,  Berger E.,
  Harrison F.,   Yost S.,  2002, GRB Coordinates Network, 1554

\bibitem[\protect\citeauthoryear{Soderberg et~al.,}{Soderberg
  et~al.}{2004}]{Soderberg04}
Soderberg A.~M.,  et~al., 2004, \mn@doi [The Astrophysical Journal]
  {10.1086/383082}, 606, 994

\bibitem[\protect\citeauthoryear{{Stanek} et~al.,}{{Stanek}
  et~al.}{2003}]{Stanek03}
{Stanek} K.~Z.,  et~al., 2003, \mn@doi [\apjl] {10.1086/376976}, \href
  {http://adsabs.harvard.edu/abs/2003ApJ...591L..17S} {591, L17}

\bibitem[\protect\citeauthoryear{Starling et~al.,}{Starling
  et~al.}{2011}]{Starling11}
Starling R. L.~C.,  et~al., 2011, \mn@doi [Monthly Notices of the Royal
  Astronomical Society] {10.1111/j.1365-2966.2010.17879.x}, 411, 2792

\bibitem[\protect\citeauthoryear{{Steeghs} et~al.,}{{Steeghs}
  et~al.}{2022}]{Steeghs22}
{Steeghs} D.,  et~al., 2022, \mn@doi [\mnras] {10.1093/mnras/stac013}, \href
  {https://ui.adsabs.harvard.edu/abs/2022MNRAS.511.2405S} {511, 2405}

\bibitem[\protect\citeauthoryear{{Steele} et~al.,}{{Steele}
  et~al.}{2004}]{Steele04}
{Steele} I.~A.,  et~al., 2004, in {Oschmann} Jacobus~M. J.,  ed.,  Society of
  Photo-Optical Instrumentation Engineers (SPIE) Conference Series Vol. 5489,
  Ground-based Telescopes. pp 679--692, \mn@doi{10.1117/12.551456}

\bibitem[\protect\citeauthoryear{Suda et~al.,}{Suda et~al.}{2021}]{Suda21}
Suda Y.,  et~al., 2021, in Proceedings of the 37th International Cosmic Ray
  Conference—PoS (ICRC2021), Berlin, Germany.

\bibitem[\protect\citeauthoryear{{Taddia} et~al.,}{{Taddia}
  et~al.}{2015}]{Taddia15}
{Taddia} F.,  et~al., 2015, \mn@doi [\aap] {10.1051/0004-6361/201423915}, \href
  {https://ui.adsabs.harvard.edu/abs/2015A&A...574A..60T} {574, A60}

\bibitem[\protect\citeauthoryear{Tang, Huang, Geng  \& Zhang}{Tang
  et~al.}{2019}]{Tang19}
Tang C.-H.,  Huang Y.-F.,  Geng J.-J.,   Zhang Z.-B.,  2019, \mn@doi [The
  Astrophysical Journal Supplement Series] {10.3847/1538-4365/ab4711}, 245, 1

\bibitem[\protect\citeauthoryear{{Tanvir}, {Levan}, {Fruchter}, {Hjorth},
  {Hounsell}, {Wiersema}  \& {Tunnicliffe}}{{Tanvir} et~al.}{2013}]{Tanvir13}
{Tanvir} N.~R.,  {Levan} A.~J.,  {Fruchter} A.~S.,  {Hjorth} J.,  {Hounsell}
  R.~A.,  {Wiersema} K.,   {Tunnicliffe} R.~L.,  2013, \mn@doi [\nat]
  {10.1038/nature12505}, \href
  {http://adsabs.harvard.edu/abs/2013Natur.500..547T} {500, 547}

\bibitem[\protect\citeauthoryear{{Tonry} et~al.,}{{Tonry}
  et~al.}{2018}]{Tonry18}
{Tonry} J.~L.,  et~al., 2018, \mn@doi [\apj] {10.3847/1538-4357/aae386}, \href
  {https://ui.adsabs.harvard.edu/abs/2018ApJ...867..105T} {867, 105}

\bibitem[\protect\citeauthoryear{{Troja} et~al.,}{{Troja}
  et~al.}{2017}]{Troja17b}
{Troja} E.,  et~al., 2017, \mn@doi [\nat] {10.1038/nature24290}, \href
  {http://adsabs.harvard.edu/abs/2017Natur.551...71T} {551, 71}

\bibitem[\protect\citeauthoryear{{Troja} et~al.,}{{Troja}
  et~al.}{2018}]{Troja18}
{Troja} E.,  et~al., 2018, \mn@doi [Nature Communications]
  {10.1038/s41467-018-06558-7}, \href
  {http://adsabs.harvard.edu/abs/2018NatCo...9.4089T} {9, 4089}

\bibitem[\protect\citeauthoryear{{Tunnicliffe} \& {Levan}}{{Tunnicliffe} \&
  {Levan}}{2012}]{Tunnicliffe12}
{Tunnicliffe} R.~L.,  {Levan} A.,  2012, in {Roming} P.,  {Kawai} N.,   {Pian}
  E.,  eds,  Proceedings IAU Symposium Vol. 279, Death of Massive Stars:
  Supernovae and Gamma-Ray Bursts. pp 415--416,
  \mn@doi{10.1017/S1743921312013610}

\bibitem[\protect\citeauthoryear{{Urata}, {Huang}, {Yamazaki}  \&
  {Sakamoto}}{{Urata} et~al.}{2015}]{Urata15}
{Urata} Y.,  {Huang} K.,  {Yamazaki} R.,   {Sakamoto} T.,  2015, \mn@doi [\apj]
  {10.1088/0004-637X/806/2/222}, \href
  {https://ui.adsabs.harvard.edu/abs/2015ApJ...806..222U} {806, 222}

\bibitem[\protect\citeauthoryear{Virgili, Liang  \& Zhang}{Virgili
  et~al.}{2008}]{Virgili08}
Virgili F.~J.,  Liang E.-W.,   Zhang B.,  2008, \mn@doi [Monthly Notices of the
  Royal Astronomical Society] {10.1111/j.1365-2966.2008.14063.x}, 392, 91

\bibitem[\protect\citeauthoryear{{White}}{{White}}{2020}]{White20}
{White} N.~E.,  2020, in Gamma-ray Bursts in the Gravitational Wave Era 2019.
  pp 51--53 (\mn@eprint {arXiv} {2003.01592})

\bibitem[\protect\citeauthoryear{{Willingale}, {Starling}, {Beardmore},
  {Tanvir}  \& {O'Brien}}{{Willingale} et~al.}{2013}]{willingale13}
{Willingale} R.,  {Starling} R.~L.~C.,  {Beardmore} A.~P.,  {Tanvir} N.~R.,
  {O'Brien} P.~T.,  2013, \mn@doi [\mnras] {10.1093/mnras/stt175}, \href
  {https://ui.adsabs.harvard.edu/abs/2013MNRAS.431..394W} {431, 394}

\bibitem[\protect\citeauthoryear{{Wilms}, {Allen}  \& {McCray}}{{Wilms}
  et~al.}{2000}]{Wilms00}
{Wilms} J.,  {Allen} A.,   {McCray} R.,  2000, \mn@doi [\apj] {10.1086/317016},
  \href {https://ui.adsabs.harvard.edu/abs/2000ApJ...542..914W} {542, 914}

\bibitem[\protect\citeauthoryear{{Yu} et~al.,}{{Yu} et~al.}{2016}]{Yu16}
{Yu} H.-F.,  et~al., 2016, \mn@doi [\aap] {10.1051/0004-6361/201527509}, \href
  {https://ui.adsabs.harvard.edu/abs/2016A&A...588A.135Y} {588, A135}

\bibitem[\protect\citeauthoryear{{Yuan} et~al.,}{{Yuan} et~al.}{2015}]{Yuan15}
{Yuan} W.,  et~al., 2015, arXiv e-prints, \href
  {https://ui.adsabs.harvard.edu/abs/2015arXiv150607735Y} {p. arXiv:1506.07735}

\bibitem[\protect\citeauthoryear{{Zhang} \& {M{\'e}sz{\'a}ros}}{{Zhang} \&
  {M{\'e}sz{\'a}ros}}{2001}]{Zhang01}
{Zhang} B.,  {M{\'e}sz{\'a}ros} P.,  2001, \mn@doi [\apjl] {10.1086/320255},
  \href {http://adsabs.harvard.edu/abs/2001ApJ...552L..35Z} {552, L35}

\bibitem[\protect\citeauthoryear{{Zhang} \& {M{\'e}sz{\'a}ros}}{{Zhang} \&
  {M{\'e}sz{\'a}ros}}{2004}]{Zhang04}
{Zhang} B.,  {M{\'e}sz{\'a}ros} P.,  2004, \mn@doi [International Journal of
  Modern Physics A] {10.1142/S0217751X0401746X}, \href
  {http://adsabs.harvard.edu/abs/2004IJMPA..19.2385Z} {19, 2385}

\bibitem[\protect\citeauthoryear{{Zhang}, {Fan}, {Dyks}, {Kobayashi},
  {M{\'e}sz{\'a}ros}, {Burrows}, {Nousek}  \& {Gehrels}}{{Zhang}
  et~al.}{2006}]{Zhang06}
{Zhang} B.,  {Fan} Y.~Z.,  {Dyks} J.,  {Kobayashi} S.,  {M{\'e}sz{\'a}ros} P.,
  {Burrows} D.~N.,  {Nousek} J.~A.,   {Gehrels} N.,  2006, \mn@doi [\apj]
  {10.1086/500723}, \href {http://adsabs.harvard.edu/abs/2006ApJ...642..354Z}
  {642, 354}

\bibitem[\protect\citeauthoryear{Zhang, Lü  \& Liang}{Zhang
  et~al.}{2016}]{Zhang16}
Zhang B.,  Lü H.-J.,   Liang E.-W.,  2016, \mn@doi [Space Science Reviews]
  {10.1007/s11214-016-0305-9}, 202

\bibitem[\protect\citeauthoryear{{Zhang}, {Zhang}, {Li}, {Su}, {Dong}, {Chang}
  \& {Zhang}}{{Zhang} et~al.}{2020a}]{Zhang20}
{Zhang} X.-L.,  {Zhang} C.-T.,  {Li} X.-J.,  {Su} F.-F.,  {Dong} X.-F.,
  {Chang} H.-Y.,   {Zhang} Z.-B.,  2020a, \mn@doi [Research in Astronomy and
  Astrophysics] {10.1088/1674-4527/20/12/201}, \href
  {https://ui.adsabs.harvard.edu/abs/2020RAA....20..201Z} {20, 201}

\bibitem[\protect\citeauthoryear{Zhang, Christie, Petropoulou, Rueda-Becerril
  \& Giannios}{Zhang et~al.}{2020b}]{Zhang20c}
Zhang H.,  Christie I.~M.,  Petropoulou M.,  Rueda-Becerril J.~M.,   Giannios
  D.,  2020b, \mn@doi [Monthly Notices of the Royal Astronomical Society]
  {10.1093/mnras/staa1583}, 496, 974

\bibitem[\protect\citeauthoryear{Zhang, Jiang, Zhang, Zhang, Li  \&
  Zhang}{Zhang et~al.}{2020c}]{Zhang20b}
Zhang Z.~B.,  Jiang M.,  Zhang Y.,  Zhang K.,  Li X.~J.,   Zhang Q.,  2020c,
  \mn@doi [The Astrophysical Journal] {10.3847/1538-4357/abb400}, 902, 40

\bibitem[\protect\citeauthoryear{Zhang et~al.,}{Zhang et~al.}{2021}]{Zhang21}
Zhang B.-B.,  et~al., 2021, \mn@doi [Nature Astronomy]
  {10.1038/s41550-021-01395-z}, 5, 911

\bibitem[\protect\citeauthoryear{{de Ugarte Postigo}, {Kann}, {Blazek}, {Agui
  Fernandez}, {Thoene}  \& {Gomez Velarde}}{{de Ugarte Postigo}
  et~al.}{2020}]{GCN28649}
{de Ugarte Postigo} A.,  {Kann} D.~A.,  {Blazek} M.,  {Agui Fernandez} J.~F.,
  {Thoene} C.,   {Gomez Velarde} G.,  2020, GRB Coordinates Network, \href
  {https://ui.adsabs.harvard.edu/abs/2020GCN.28649....1D} {28649, 1}

\makeatother
\end{thebibliography}

%%%%%%%%%%%%%%%%%%%%%%%%%%%%%%%%%%%%%%%%%%%%%%%%%%

%%%%%%%%%%%%%%%%% APPENDICES %%%%%%%%%%%%%%%%%%%%%
\newpage
\section{Appendices}

%\subsection{BAT and XRT fitting results}
%In Table~\ref{tab:modelfit} we present the results of fitting models to the BAT and XRT spectra of GRB 201015A. The description of the model and parameters can be found in the HEASARC documentation for {\sc xspec}.

%\textbf{
The optical afterglow and supernova observations plotted in Figure~\ref{fig:Afterglow_lc} are given in Table S1 of the Supplementary material, and a sample of the table is shown below.%}
These data were used for the power-law / broken power-law, and SN model fitting. The results from the fitting were used to determine the spectral and temporal indices, deceleration time of the afterglow, and SN peak time and luminosity.

%\subsection{Optical afterglow data}
The optical afterglow and supernova observations plotted in Figure~\ref{fig:Afterglow_lc} are given in Table 9. These data were used for the power-law/ broken power-law, and SN model fitting. The results from the fitting were used to determine the spectral and temporal indices, deceleration time of the afterglow, and SN peak time and luminosity.

\begin{table}
\begin{tabular}{|c|c|c|c|c|c|c|}
\hline 
time(s)&observed magnitude &err low &err high &filter&telescope&GCN\\
\hline
73.2&17.67&0.1&0.1&i'&NUTTelA-TAO/BSTI&28674\\
73.2&17.79&0.1&0.1&r'&NUTTelA-TAO/BSTI&28674\\
98.0&18.34&0.08&0.08&r&NEXT-0.6m&28653\\
103.2&17.86&0.1&0.1&i'&NUTTelA-TAO/BSTI&28674\\
103.2&18.16&0.1&0.1&r'&NUTTelA-TAO/BSTI&28674\\
157.6&16.96&0.1&0.1&i'&NUTTelA-TAO/BSTI&28674\\
157.6&17.91&0.15&0.15&g'&NUTTelA-TAO/BSTI&28674\\
157.6&16.81&0.1&0.1&r'&NUTTelA-TAO/BSTI&28674\\
202.6&16.75&0.1&0.1&i'&NUTTelA-TAO/BSTI&28674\\
202.6&17.76&0.15&0.15&g'&NUTTelA-TAO/ BSTI&28674\\
\hline
\end{tabular}
\label{tab:optical}
\caption{%\textbf{
This table lists a sample of the optical data on GRB 201015A used for analysis in this paper. The rest of the data can be found in the Supplementary material.}
%}
\end{table}

%%%%%%%%%%%%%%%%%%%%%%%%%%%%%%%%%%%%%%%%%%%%%%%%%%

% Don't change these lines
\bsp	% typesetting comment
\label{lastpage}
\end{document}